\documentclass{aa}

\usepackage{graphicx}
\usepackage{txfonts}
\usepackage{siunitx}
\usepackage{multirow}
\usepackage{color}
\def\Mjup{\hbox{$\thinspace M_{\mathrm{J}}$}}
\def\Msun{\hbox{$\thinspace M_{\odot}$}}
\def\Rsun{\hbox{$\thinspace R_{\odot}$}}
\def\Lsun{\hbox{$\thinspace L_{\odot}$}}
\def\Teff{\hbox{$\thinspace T_{\mathrm{eff}}$}}

\def\kms{\hbox{$\thinspace {\mathrm{km~s^{-1}}}$}}
\def\ms{\hbox{$\thinspace {\mathrm{m~s^{-1}}}$}}

\def\ALi{\hbox{$\thinspace A (\mathrm{Li})$}}

\def\au{\hbox{$\thinspace \mathrm{au}$}}
\def\sjit{\hbox{$\sigma_{\mathrm{jitter}}$}}

\def\Prot{\hbox{$P_{\mathrm{rot}}$}}

\def\ccr{\hbox{$\thinspace ^{12}\mathrm{C}/^{13}\mathrm{C}$}} 
\def\shk{\hbox{$S_{\mathrm{HK}}$}}
\def\ha{\hbox{$I_{\mathrm{H_{\alpha}}}$}}
\def\ife{\hbox{$I_{\mathrm{Fe}}$}}

\def\starA{BD+48~740\thinspace}               
\def\starB{HD~107028\thinspace}              
\def\starC{TYC~0684-00553-1\thinspace} 
\def\starD{TYC~3105-00152-1\thinspace} 
\def\starE{TYC~3263-02180-1\thinspace} 
\def\starF{HD~181368\thinspace}              
\def\starG{BD+02~3497\thinspace}            
\def\starH{HD~188214\thinspace}             
\def\starI{TYC~3300-00133-1\thinspace}  
\def\starJ{TYC~3318-01333-1\thinspace} 
\def\starK{TYC~3663-01966-1\thinspace} 
\def\starL{TYC~3930-00681-1\thinspace} 
\def\starM{HD~238914\thinspace}            

\begin{document} 

  \title{Tracking Advanced Planetary Systems (TAPAS) with HARPS-N.  
  \thanks{Based on observations obtained with the Hobby-Eberly Telescope, 
  which is a joint project of the University of Texas at Austin, the Pennsylvania State University, 
  Stanford University, Ludwig-Maximilians-Universit\"at M\"unchen, and Georg-August-Universit\"at G\"ottingen.}
  \thanks{Based on observations made with the Italian Telescopio Nazionale Galileo (TNG) operated 
  on the island of La Palma by the Fundaci\'on Galileo Galilei of the INAF (Istituto Nazionale di Astrofisica) 
  at the Spanish Observatorio del Roque de los Muchachos of the Instituto de Astrof\'{\i}sica de Canarias.}
  \thanks{Table 2 is only available in electronic form
at the CDS via anonymous ftp to cdsarc.u-strasbg.fr (130.79.128.5)
or via http://cdsweb.u-strasbg.fr/cgi-bin/qcat?J/A+A/}
}

   \subtitle{VI.  \starM  and \starJ  - two more Li-rich giants with planets}

   \titlerunning{TAPAS VI. Two more Li-rich giants with planets}
   \authorrunning{M. Adam\'ow et al.}

 \author{M. Adam\'ow
          \inst{1,2}
         \and 
         A. Niedzielski
          \inst{2}  
             \and
          K. Kowalik
          \inst{3}         
          \and                            
          E. Villaver
         \inst{4}
          \and                                                          
          A. Wolszczan
          \inst{5,6}
          \and                                
         G. Maciejewski
          \inst{2}          
          \and
          M. Gromadzki
          \inst{7}
                  }
                   
   \institute{ McDonald Observatory and Department of Astronomy, University of Texas at Austin, 2515 Speedway, Stop C1402, Austin, Texas, 78712-1206, USA.
   \email{madamow@icloud.com}
   \and   
   Toru\'n Centre for Astronomy, Faculty of Physics, Astronomy and Applied Informatics, Nicolaus Copernicus University in Toru\'n, Grudziadzka 5, 87-100 Toru\'n, Poland.
   \email{Andrzej.Niedzielski@umk.pl}
   \and
   National Center for Supercomputing Applications, University of Illinois, Urbana-Champaign, 1205 W Clark St, MC-257, Urbana, IL 61801, USA                  
    \and
    Departamento de F\'{\i}sica Te\'orica, Universidad Aut\'onoma de Madrid, Cantoblanco 28049 Madrid, Spain.
    \email{Eva.Villaver@uam.es}
   \and
    Department of Astronomy and Astrophysics, Pennsylvania State University, 525 Davey Laboratory, University Park, PA 16802, USA
   \and
    Center for Exoplanets and Habitable Worlds, Pennsylvania State University, 525 Davey Laboratory, University Park, PA 16802, USA
    \and 
    Warsaw University Astronomical Observatory, Al. Ujazdowskie 4, PL-00-478, Warszawa, Poland.
     }

   \date{Received;accepted}

 
  \abstract
   { We present the latest results of our search for planets with  HARPS-N at the 3.6 m Telescopio Nazionale Galileo
   under the Tracking Advanced Planetary Systems project: an in-depth study of the 15 most Li abundant giants 
   from the PennState - Toru\'n Planet Search sample. }
   {Our goals are first, to obtain radial velocities of the most Li-rich giants we identified in our sample 
   to search for possible low-mass substellar companions, and second, to perform an extended spectral analysis to
   define the evolutionary status of these stars. }
   {This work is based on high-resolution spectra obtained with the Hobby-Eberly Telescope 
   and its High Resolution Spectrograph, and with the HARPS-N spectrograph at the Telescopio Nazionale Galileo. 
   Two stars, \starF and \starH, were also observed with UVES at the VLT to determine beryllium abundances.}
   {We report i)  the discovery of two new planetary systems around the Li-rich giant stars: \starM  and \starJ (a binary system); 
   ii) reveal a binary Li-rich giant,  \starF; 
   iii) although our current phase coverage is not complete, we suggest the presence of planetary mass companions around \starK and \starD; 
   iv) we confirm the previous result for \starA\ and present updated orbital parameters, and 
   v) we find a lack of a relation between the Li enhancement and the Be abundance for 
   the stars \starF and \starH, for which we acquired blue spectra.}
   {We found seven stars with stellar or potential planetary companions among the 15 Li-rich giant stars. 
   The binary star frequency of the Li-rich giants in our sample appears to be normal, but the planet frequency 
   is twice that of the general sample, which suggests a possible connection between hosting a companion 
   and enhanced Li abundance in giant stars. We also found most of the companions orbits to be highly eccentric. }


   \maketitle
%

\section{Introduction}
The hypothesis that a star can enhance its lithium content in the atmosphere through planet engulfment has been formulated by \cite{Alexander1967}
long before any Li-rich giant (star)  or a planetary host was discovered.

The first Li-rich giant was discovered 25 years after Alexander's work by \cite{WallersteinSneden1982}.
Today, we know more than 150  stars with $\ALi>1.5$. They are  most often defined  as Li-rich objects,
but it is still not clear  why one giant star in 100 objects has an enhanced Li abundance. The variety of
properties of Li-rich giants is intriguing: many are located all along the red giant branch (RGB) \citep{Lebzelter2012},
and some are identified as stars at the turn-off point \citep{Koch2011} and horizontal branch objects. Furthermore, 
for some stars, the high Li content is associated with infrared excesses \citep{Kumar2015,Rebull2015}. 

The first candidate for a Li-rich giant hosting a planet was reported by \cite{Adamow2012}
as a result of the PennState - Toru\'n Planet Search (PTPS). Currently, we know only a few more Li-rich objects that have
been identified as potential
planetary hosts: 8~UMi (HD~133086) \citep{Kumar2011, Lee2015}, NGC~2423~3 \citep{Carlberg2016},
and two giants in the NGC~2423 and  NGC~4349 clusters \citep{DelgadoMena2016}.

Under special circumstances (efficient mixing between the convective layer and the hydrogen burning shell), 
a chain of chemical reactions that lead to Li production can occur \citep{CameronFowler1971}, 
most likely at the luminosity function bump in the Herztsprung-Russel
(HR) diagram. This means that an RGB  star
may be able to produce Li by itself (\citealt{Adamow2014}  and references therein).
Li enhancement may be also the result of a transfer of Li-rich gas from a more evolved companion
in a close binary system \citep{SacBoo1999}. Rapid Li production is also associated with 
supernova (SN) explosions  \citep{WoosleyWeaver1995}. If a star accretes post-SN material, 
its chemical composition may be altered, resulting in higher Li content.
Planetary ingestion is
expected when a star leaves the main sequence and moves to the tip of the RGB,
where the significant expansion in radius and the extension of the convective zone amplifies the effects of tidal interactions 
with the orbiting planets \citep{Villaver2009}. The lack of close-in planets around evolved stars suggests that planetary engulfment
is a common process in the evolution of planetary systems, and the solar system probably is no exception \citep{Villaver2009, Villaver2014}.

The increasing number of evolved stars hosting planets with enhanced Li abundances may
suggest that indeed planet engulfment is the mechanism responsible for Li abundance anomalies. 
The amount of unprocessed matter required to enhance the Li abundance in a giant is very high.  
However, the hypothesis that is more attractive to study in more detail are the structural 
effects on the star and the chemical  signatures  of  a planet engulfment. \cite{PriviteraI} showed 
that planet engulfment either produces a too weak signal, as in the case 
of the carbon isotopic ratios, or produces signals that cannot be attributed
in a non-ambiguous way to a planet engulfment.

It is very tempting, however,
to search for low-mass companions around Li-rich giants since a large portion of planetary systems are expected to be multiple.
PTPS, a planet-search program focused on 
evolved stars, is the perfect laboratory in which to study this problem in detail. 
Within the project, we collected high-resolution spectra that we used to 
determine Li abundances for the whole sample of nearly 1000 objects, which are mostly composed of evolved stars. 
We identified several new Li-rich stars \citep{Adamow2012MSAIS,Adamow2014} for which 
we decided to perform a dedicated campaign to investigate their
properties,
including  radial velocity (RV) variations  and Be abundances, as a part of the
Tracking Advanced Planetary Systems (TAPAS) project \citep{TAPAS1, TAPAS2,TAPAS3,TAPAS4,TAPAS5},
 to test the working hypothesis that  Li-rich giants might be related to the process 
 of planet engulfment and might even still host more planets.

In this paper, we present results of RV measurements for 12 stars 
with the highest Li abundances identified in the PTPS sample ($\ALi_{\mathrm{NLTE}}\geq1.4$),
and we give an update on orbital parameters of \starA~b after numerous additional epochs of observations.
We also present results of chemical composition studies for those stars,
including Be abundances in two of them: \starF and \starH, which
are bright enough and available for high-resolution spectroscopy in the UV. 


\section{Sample and observations\label{observations}}

The sample presented in this paper comprises a total of 15 Li-rich giants from PTPS
with the  highest Li content, $\ALi_{\mathrm{NLTE}}\geq1.4$ \citep{Adamow2014}. 
\starF was added to this list because this star was identified as a giant by \cite{Deka2017}, and it was found to be a Li-rich giant as well.
Twelve objects are the subject of detailed analysis, as new epochs 
of precise RV were collected in a quest for possible companions around them within the TAPAS project.
The sample  also includes \starA \citep{Adamow2012}, for which new RV observations are available.

We also added three stars to this sample that had no new observation
epochs: 
the apparently single Li-rich giant  with the highest A(Li) in our sample, 
\starB \citep{TAPAS2},
and two Li-rich giants in binary systems, TYC~0405-01700-1 and TYC~3314-01371-1, which were presented in \cite{Adamow2014} 
for completeness. 

Basic atmospheric parameters ($\Teff$, $\log g$, $\text{and }[$Fe/H$]$, determined through
abundance analysis of neutral and ionized Fe lines) and Li abundance
are presented in Table \ref{parameters}. A detailed description of the atmospheric parameters can be found in 
\cite{Zielinski2012} and \cite{NiedzielskiDeka2016}. Masses, radii, and stellar ages were adopted from \cite{Deka2017}.

\begin{table*}
\centering
\tiny
\caption{Parameters of PTPS stars with high Li abundances. }
  \begin{tabular}{l | S[table-format=4.0(2)] S[table-format=1.2(3)] 
                      S[table-format=1.2(3)] S[table-format=1.2(3)] 
                      S[table-format=1.2(3)] S[table-format=2.2(3)] 
                      S[table-format=2.2(3)] S[table-format=4.0(4)]
                      S[table-format=1.2]}
\hline
    Star &  {$\Teff$[K]} & {$\log g$}  & {$[$Fe/H$]$}   & {$\log L/\Lsun$}  & {$M/\Msun$}     & {$R/\Rsun$}     & {$v
    \sin i\;[\!\kms] $} & {$\Prot\;[\mathrm{days}]$} & {$\ALi$} \\
\hline \hline
\starG & 4955 \pm 8  & 3.03 \pm 0.03 & -0.05 \pm 0.05 & 1.47 \pm 0.15 & 1.81 \pm 0.24 &  7.08 \pm 1.23 & 0.70 \pm 3.60 &  511 \pm 2720  & 1.60 \\ 
\starA & 4534 \pm 8  & 2.48 \pm 0.04 & -0.13 \pm 0.06 & 1.64 \pm 0.12 & 1.09 \pm 0.16 & 10.33 \pm 1.81 & 0.70 \pm 2.20 &  746 \pm 2477  & 2.07 \\ 
\starF & 4852 \pm 15 & 2.49 \pm 0.05 & -0.06 \pm 0.02 & 0.83 \pm 0.10 & 0.96 \pm 0.13 &  6.45 \pm 0.79 & 2.43 \pm 0.97 &  134 \pm 70    & 2.07 \\ 
\starH & 4758 \pm 13 & 3.04 \pm 0.05 & -0.17 \pm 0.07 & 1.79 \pm 0.18 & 1.17 \pm 0.08 &  8.50 \pm 1.48 & 6.00 \pm 0.40 &   71 \pm 17    & 1.44 \\
\starM & 4769 \pm 15 & 2.37 \pm 0.06 & -0.25 \pm 0.09 & 1.85 \pm 0.19 & 1.47 \pm 0.47 & 12.73 \pm 3.89 & 2.50 \pm 0.50 &  257 \pm 130   & 1.99  \\
\starC & 4719 \pm 10 & 2.38 \pm 0.03 & -0.18 \pm 0.05 & 1.90 \pm 0.14 & 1.61 \pm 0.35 & 13.42 \pm 2.57 & 0.60 \pm 2.80 & 1131 \pm 5497  & 2.92 \\ 
\starD & 4673 \pm 5  & 2.45 \pm 0.02 & -0.14 \pm 0.05 & 1.66 \pm 0.04 & 1.22 \pm 0.08 & 10.62 \pm 0.83 & 1.90 \pm 0.60 &  282 \pm 111   & 2.86 \\ 
\starE & 4915 \pm 10 & 2.27 \pm 0.04 & -0.72 \pm 0.01 & 1.74 \pm 0.12 & 1.02 \pm 0.12 & 11.25 \pm 1.97 & 3.09 \pm 0.94 &  184 \pm 88    & 1.92 \\ 
\starI  & 5007 \pm 10 & 3.08 \pm 0.03 &  0.17 \pm 0.06 & 1.41 \pm 0.10 & 2.04 \pm 0.16 &  6.76 \pm 0.92 & 3.00 \pm 0.70 &  114 \pm 42    & 1.55 \\ 
\starJ & 4776 \pm 10 & 2.97 \pm 0.04 & -0.06 \pm 0.06 & 1.21 \pm 0.11 & 1.19 \pm 0.14 &  5.90 \pm 1.00 & 1.50 \pm 0.70 &  198 \pm 126   & 1.51 \\ 
\starK & 5068 \pm 10 & 2.40 \pm 0.03 & -0.26 \pm 0.05 & 2.26 \pm 0.09 & 2.88 \pm 0.21 & 17.66 \pm 2.28 & 2.70 \pm 0.60 &  330 \pm 116   & 1.41 \\ 
\starL & 4777 \pm 20 & 2.89 \pm 0.08 & -0.10 \pm 0.09 & 1.38 \pm 0.23 & 1.25 \pm 0.26 &  6.90 \pm 2.29 & 2.20 \pm 0.60 &  158 \pm 95    & 1.42 \\ 
\hline
\starB & 5171 \pm 15 & 3.16 \pm 0.04 & -0.09 \pm 0.02 & 1.44 \pm 0.08 & 1.85 \pm 0.05 &  6.22 \pm 0.51 & 1.46 \pm 0.44 &  215 \pm 82    & 3.65 \\ 
TYC~0405-01700-1 & 4626 \pm 50 & 2.79 \pm 0.17 & -0.15          & 1.52          & 1.4           & 9.1            & 18.9 \pm 2.1  &   24           & 1.50 \\
TYC~3314-01371-1 & 5019 \pm 10 & 3.25 \pm 0.04 & -0.06 \pm 0.07 & 1.02 \pm 0.11 & 1.30 \pm 0.16 &  4.39 \pm 0.75 & 2.00 \pm 0.80 &  111 \pm 63    & 1.45 \\
\hline
\end{tabular}
\label{parameters}
\end{table*}

\begin{figure}
   \centering
   \includegraphics[width=0.5\textwidth]{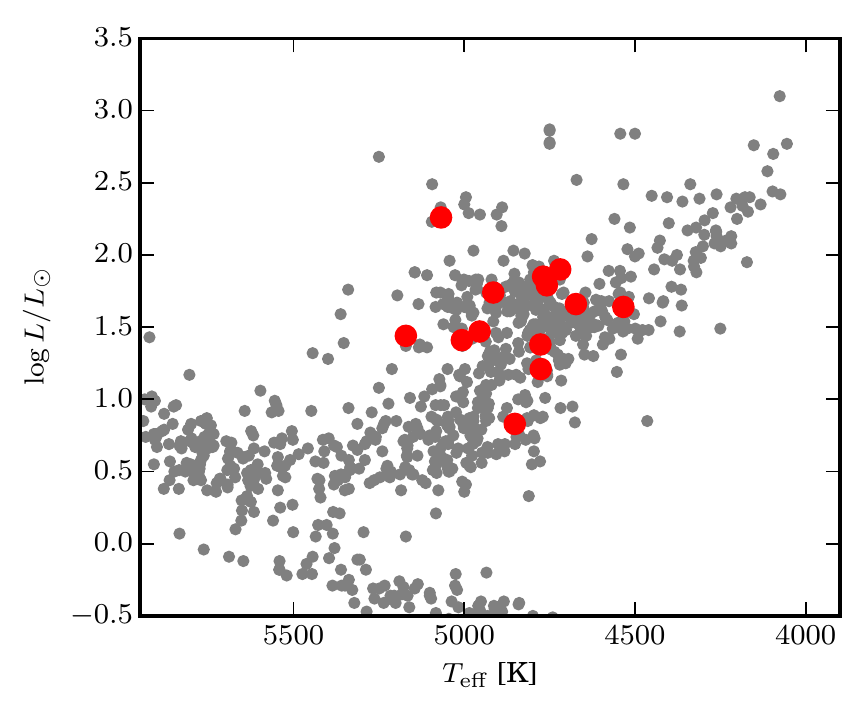}
   \caption{The Hertzsprung-Russell diagram for all PTPS stars (gray). The red points represents stars analyzed in this work.}
   \label{HR}
\end{figure} 

The spectroscopic observations presented in this paper were made  with
the 9.2  Hobby-Eberly Telescope
(HET, \citealt{Ramsey1998}) and its  High-Resolution Spectrograph (HRS,
\citealt{Tull1998}) in the queue-scheduled mode \citep{Shetrone2007},
and with the 3.58 meter Telescopio Nazionale Galileo (TNG) and its High
Accuracy Radial velocity Planet Searcher in the North hemisphere (HARPS-N,
\citealt{Cosentino2012}).
A detailed description of the adopted observing strategies and instrumental configuration for
HET/HRS and TNG/HARPS-N can be found in  \cite{Niedzielski2007} and \cite{TAPAS1}.

 Additionally, we obtained UVES \citep{UVES2000} spectra for two stars:
\starF and \starH. The configuration of the spectrograph was chosen
to cover the bluest spectral region available for UVES, which includes the beryllium resonance lines.
Five exposures were made for \starF, and six for \starH in 2014 (April 4 and 10; June 28; and July 2 and 10).
All observations were reduced with {\it Reflex}  - a dedicated reduction pipeline \citep{Reflex}. Single exposures 
for each star were combined to achieve a spectrum with a  high signal-to-noise ratio, which we used for
determining the abundance of beryllium.

All HET/HRS spectra were reduced with standard IRAF\footnote{IRAF is distributed 
by the National Optical Astronomy Observatories, which are operated by the Association 
of Universities for Research in Astronomy, Inc., under cooperative agreement with
the National Science Foundation.} procedures.
The TNG/HARPS-N spectra were processed with the standard user's pipeline. The typical signal-to-noise ratio
for these spectra is 50-80. As HARPS-N is a  temperature- and pressure-controlled 
spectrograph,  all gathered spectra for a star were combined to obtain one high-resolution
spectrum with a  high signal-to-noise ratio for the purpose of the chemical analysis.

\section{Radial velocities and activity\label{RV}}

Stellar activity may be a source of RV-like variations in stars. Spots on the stellar surface, which move across the disk as the star rotates,
distort the line shapes and influence the RV measurements. 
A long-term stellar magnetic cycle may possibly also influence the RV measurements through a convective blueshift. This effect has been intensively studied 
for main-sequence stars (\citealt{Meunier2017a, Meunier2017b} and references therein).
For a proper identification of the RV variability in a source, we
investigated four indicators of stellar activity. 

\subsection{Radial velocities}
All HET/HRS RV and spectral line bisector (BIS) measurements  were obtained using the I$_2$ gas-cell technique  \citep{MarcyButler1992, Butler1996}. 
Our application of this technique to HET/HRS data is described in detail in \cite{Nowak2012} and \cite{ Nowak2013}. 

The RV and BIS from HARPS-N were obtained with the cross-correlation \citep{Queloz1995, Pepe2002} method,
with the standard user's pipeline, which
uses  the simultaneous Th-Ar calibration mode of the spectrograph  and  the cross-correlation  
mask with stellar spectrum (mask). For all stars, except for \starB and~\starE, 
we used the K5 mask. For these two stars, we used the G2 mask.
The RV and BIS data for all stars are presented in Table 2.

The RVs discussed in this paper were collected in a long period of time -- HET/HRS data were collected between 2 June 2004 and 14 July 2015, 
TNG/HARPN-N were obtained between 30 November 2012 and 4 August 2015. 
The time span of observations for each star is given in Table \ref{RV-summary}.
This table also includes the estimated amplitudes of solar-like oscillations $K_{\mathrm{osc}}$ \citep{1995A&A...293...87K}, 
associated with granulation and other heterogeneities on the
stellar surface,
the initial amplitudes in  RV (defined as the difference between 
maximum and minimum values of raw RVs), the average error in RV,
the linear correlation coefficient between RV and BIS, and its significance level. 

Table \ref{RV-summary}  shows that the observed RV variations are 1-2 orders 
of magnitude larger than the average RV uncertainty for all our
stars.  In most cases, the observed RV variations 
are at least an order of magnitude larger than the expected RV amplitude of the p-mode oscillations. 
The only exceptions are \starH - whose HET/HRS  RV amplitude is only twice larger 
than the expected $K_{\mathrm{osc}}$, not confirmed by TNG/HARPS-N data (therefore most likely 
caused by the low number of HET/HRS epochs), and \starL, again most likely due to 
the low number of HET/HRS observations.

There is no statistically significant correlation between RV and BIS in HET/HRS data
(except for \starL, for which there are just  two observation
epochs)
In the TNG data for two stars, \starG and  \starE,  the linear correlation coefficients
are below a level of significance of $\mathrm{p}<0.05$, pointing to a relation between RV and BIS.
For other stars we can assume that the RV signal origin is Dopplerian.

\setcounter{table}{2}
\begin{table*}
\centering
  \caption{Summary of RV observations. $OP_{\mathrm{total}}$ denotes the
  total observation period in days, $OP_{\mathrm{HET}},\;OP_{\mathrm{TNG}}$ are observing periods for the HET and TNG, respectively, $K_{\mathrm{osc}}$ is the amplitude of
  solar-like oscillations \citep{1995A&A...293...87K} in $\ms$, $K$ denotes the
  amplitude of the observed radial velocities defined as $RV_{\mathrm{max}} -RV_{\mathrm{min}} $, in $\ms$,
  $\overline{\sigma_{\mathrm{RV}}}$ is the average RV uncertainty in $\ms$, and
  $r_{\mathrm{BIS}}$ is the linear correlation coefficient between the RV and BIS.
  }
\begin{tabular}{l | c S[table-format=2.2] |
    S[table-format=4.1] S[table-format=2.1] S[table-format=1.1] S[table-format=1.2] S[table-format=1.2] r |
    S[table-format=4.1] S[table-format=2.1] S[table-format=1.1] S[table-format=1.2] S[table-format=1.2] r}
\hline 
\multirow{2}{*}{Star} & & &\multicolumn{6}{c|}{HET/HRS}& \multicolumn{6}{c}{TNG/HARPS-N}\\
  &$OP_{\mathrm{total}}$ & $K_{\mathrm{osc}}$ &
  $OP_{\mathrm{HET}}$ & $K$&$\overline{\sigma_{\mathrm{RV}}}$ & $r_{\mathrm{BIS}}$  & p&no &
   $OP_{\mathrm{TNG}}$ &$K$ & $\overline{\sigma_{\mathrm{RV}}}$ & $r_{\mathrm{BIS}}$ &p& no\\

\hline   
\starG & 3693 &  3.82 &  2825 &    72.4 & 7.0 & -0.12 &  0.70 & 12 & 827 &     65.2 &  2.2 & -0.47 & 0.05 & 18 \\ 
\starA  & 3492 &  9.37 &  2594 &   116.0 & 5.5 &  0.11 &  0.61 & 25 & 962 &   163.5 &  1.4 &  0.33 & 0.21 & 16 \\ 
\starF  & 2981 &  1.65 &  2221 & 3908.4 & 6.1 & -0.13 & 0.73 &   9 & 769 & 1776.4 &  1.6 &  0.16 & 0.53 & 18 \\ 
\starH & 3333 & 12.33 & 1873 &     24.8 & 5.0 & -0.16 & 0.84 &   4 & 977 &     82.4 &  2.6 & -0.15 & 0.62 & 14 \\
\starM & 3944 & 11.27 & 2935 &   121.3 & 5.8 & -0.15 & 0.20 & 71 & 769 &     98.7 &  1.2 &  0.33 & 0.23 & 15 \\
\starC & 2340 & 11.54 & 1588 &   138.0 & 7.2 &  0.53 & 0.14 &   9 & 834 &     91.3 &  3.4 &  0.39 & 0.11 & 18 \\ 
\starD & 2868 &  8.77 & 2065  &  137.2 &  5.7 & -0.12 & 0.70 & 14 & 827&    123.2 &  2.3 &  0.21 & 0.36 & 22 \\ 
\starE & 3320 & 12.61 & 2408 &  131.7 &  8.9 & 0.15  & 0.67 & 10 & 977 &   137.8 &  1.6 &  0.47 & 0.04 & 19 \\ 
\starI & 3940 &   2.95 & 1742 &    56.1 &  6.2 & 0.06  & 0.96 &   3 & 592 &     45.1 &  3.3 & -0.14 & 0.68 & 11 \\ 
\starJ & 3514 &   3.19 & 2630 &  306.0 &  5.5 & 0.46  & 0.14 & 12 & 962 &   241.8 &  2.2 & -0.40 & 0.12 & 16 \\ 
\starK & 2940 & 14.70 & 2500&     91.7 &  7.7 & -0.32 & 0.16 & 20 & 650 &     67.7 &  3.2 & -0.25 & 0.43 & 12 \\ 
\starL & 3769 &   4.49 & 1068&       7.2 &  5.5 & 1.00  & 0.00 &   2 & 496 &     80.0 &  3.1 &  0.34 & 0.30 & 11  \\ 
\hline
\end{tabular}
\label{RV-summary}
\end{table*}

\subsection{ $\ha$ activity index}

The core of the H$_{\alpha}$ line is an indicator of stellar activity. Changes in flux in the line center can be quantified with  
the $\ha$ activity index. It specifies the flux ratio at the core of H$_{\alpha}$ to the flux in the continuum close to this line. 
 We  measured the H$_{\alpha}$ activity index ($\ha$) in both HET/HRS and TNG/HARPS-N spectra. 
 For the calculations, we  followed the procedure described  by \cite{2013AJ....146..147M}, which was based on the approach 
 presented by \cite{2012A&A...541A...9G} and \citet[][and references therein]{2013ApJ...764....3R}. 
 The analysis of $\ha$ for HET/HRS had to be extended because of the use of $\mathrm{I_{2}}$ cell, therefore 
 the order of spectra with H$_{\alpha}$  is affected by weak $\mathrm{I_{2}}$ lines.
Moreover, the instrumental profile for HET/HRS data may vary. 
To take these two effects into account, we determined the analogical index to $\ha$  for
 Fe~I 6593.883~{\AA}.  This line is located in the same echelle order of spectra 
as H$_{\alpha}$ in our HET/HRS data and is not affected by stellar activity.
We will denote the Fe activity index as $\ife$.
If there is a correlation between RV and $\ife$ , it means that the instrumental profile strongly affects the analysis.
We also performed the same analysis on the iodine flat-field spectra (Flat GC),
to determine whether the  potential $\ha$ variability might stem from $\mathrm{I_{2}}$ itself.
Any correlation between indexes and the RV measured for Flat GCs and RVs indicates that the influence
of weak iodine lines is not negligible and the analysis may not be reliable.
The HARPS-N is an instrument with a stable instrumental profile, but we continued to  monitor the control line.
The resulting correlation coefficients and their significance levels are presented in Tables \ref{iha_het} and \ref{iha_tng}.

The HET/HRS data show statistically significant correlations between  $\ha$ 
and RV for \starC. A strong correlation between BIS and $\ife$ is present for \starA, which is mirrored in the
Flat GC
analysis. This strong correlation is also visible for $\ha$ vs. BIS in Flat GC for \starE.
 Weak iodine lines affect the analysis of these two objects.

The three stars \starM, \starE and \starI show a statistically 
significant correlation between $\ha$ and RV in the HARPS-N spectra. For the first object,  \starM, the correlation is present for all
investigated parameter pairs in the $\ha$ and $\ife$ analysis. Because the chosen control  line is not sensitive to chromospheric activity, we 
claim that these correlations have a non-stellar source.
 \starI  also shows a high, but not statistically significant correlation between RV 
and the control Fe line index, which is a strong suggestion that
the nature of the observed $\ha$ and RV correlation is instrumental.
In the absence of a similar correlation in the control Fe line and taking into account the increased (but not statistically significant) 
correlation between $\ha$ and BIS,  we assume for \starE
that these results are an indication of stellar activity.

Statistically significant correlations are also present for \starC: BIS seems to be correlated with $\ha$, but the same 
correlation is present for $\ife$, thus it is most likely caused by instrumental effects. \starJ shows a weak correlation between BIS and $\ha$.
For two objects, \starF and \starD, we detected correlations in the control line, but not in H$_\alpha$.

\begin{table*}
\centering

\caption{Linear correlation coefficients calculated for $\ha$ and the control line index  $I_{\mathrm{Fe}}$ vs. the RV and BIS data sets.  
One observational epoch for \starD was removed because there are no data for BIS.}
\tiny
\begin{tabular}{
  l |
  S[table-format=1.2] | S[table-format=1.2] | S[table-format=1.2] | S[table-format=1.2] | 
  S[table-format=1.2] | S[table-format=1.2] | S[table-format=1.2] | S[table-format=1.2] | 
  S[table-format=1.2] | S[table-format=1.2] | S[table-format=1.2] | S[table-format=1.2] | 
  S[table-format=1.2] | S[table-format=1.2] | S[table-format=1.2] | S[table-format=1.2] | 
  r}
\hline
\multirow{4}{*}{Star}    &\multicolumn{8}{c|}{HET/HRS star} & \multicolumn{8}{c|}{HET/HRS flat GC}   \\
 &\multicolumn{4}{c|}{$\ha$} &\multicolumn{4}{c|}{$\ife$} & \multicolumn{4}{c|}{$\ha$} &\multicolumn{4}{c|}{$\ife$} &   \\
  &\multicolumn{2}{c|}{RV} &\multicolumn{2}{c|}{BIS}&\multicolumn{2}{c|}{RV}& \multicolumn{2}{c|}{BIS}&
    \multicolumn{2}{c|}{RV} &\multicolumn{2}{c|}{BIS}&\multicolumn{2}{c|}{RV}& \multicolumn{2}{c|}{BIS}& no \\
    &r &p&r &p&r &p&r &p&r &p&r &p&r &p&r &p\\
  \hline
  \starG &  0.19 & 0.55 & 0.20 & 0.53 &  0.08 & 0.81 & 0.12 & 0.71 &  0.18 & 0.57 & 0.08 & 0.81 &  0.09 & 0.78 &  0.08 & 0.79 & 12 \\ 
  \starA &  0.15 & 0.47 &-0.10 & 0.64 &  0.07 & 0.75 & 0.44 & 0.03 &  0.10 & 0.63 &-0.07 & 0.73 &  0.07 & 0.75 &  0.43 & 0.03 & 25 \\ 
  \starF & -0.09 & 0.82 & -0.25 & 0.51 & -0.07 & 0.85 & -0.48 & 0.19 & 0.04 &0.91 &-0.61 & 0.08 & -0.06 & 0.88 & -0.49 & 0.18 &  9 \\ 
  \starH & -0.31 & 0.69 & -0.87 & 0.13 & -0.95 & 0.05 &  0.41 & 0.59 & -0.40 & 0.60 &-0.57 &0.43 & -0.76 &0.24 &  0.72 &0.28 &  4 \\
  \starM &  0.03 & 0.79 & 0.09 & 0.45 &  0.04 & 0.73 &   0.11 &0.37 & -0.01 &0.96 & 0.05 &0.67 & -0.06 & 0.59 &  0.07 & 0.59 & 71\\
  \starC &  0.68 & 0.04 & 0.51 & 0.16&  0.50 & 0.17 &  0.51 & 0.16 &  0.61 &0.08 & 0.60 &0.09 &   0.51& 0.16 &  0.51 & 0.16 &  9 \\ 
  \starD &  0.03 & 0.91 &-0.36 &0.22& -0.11 &0.70 &  0.38 &0.20 &  0.07 &0.82 &-0.34 & 0.25& -0.14 &0.63 &  0.38 &0.20& 13 \\ 
  \starE & -0.33 & 0.36 & 0.53 & 0.12 & 0.05 & 0.89 & -0.41 & 0.24 & -0.17 & 0.64 &  0.75 & 0.01 &  0.08 & 0.82 & -0.37 & 0.29 & 10 \\ 
  \starI & -0.91 & 0.27 & 0.35 & 0.77 & -0.07 & 0.96 &  0.99& 0.08 & -0.95 & 0.21 & 0.27 & 0.83 & -0.45 & 0.71 &  0.87 & 0.33 &  3 \\ 
  \starJ & -0.05 & 0.88 & 0.36 & 0.25 & -0.39 &0.21 &  0.09 &0.77 & -0.09& 0.78 & 0.24 & 0.45 & -0.39 &0.21 &  0.10 &0.75 & 12 \\ 
  \starK &  0.10 & 0.67& 0.07 & 0.77 & -0.01 & 0.98 &  0.30 &0.20 &  0.02 & 0.94 & 0.13 &0.58 & -0.01 &0.95 &   0.31 &0.18 & 20 \\ 
  \starL &  1.00 &0.00 & 1.00 &0.00 &  1.00 &0.00 &  1.00& 0.00 &  1.00 &0.00 & 1.00 &0.00 &  1.00&0.00 &   1.00&0.00 &  2\\ 

  \hline
  \end{tabular}
  \label{iha_het}
\end{table*}

\begin{table*}
\centering
\caption{Linear correlation coefficients calculated for $\ha$ and the control line index  $I_{\mathrm{Fe}}$ vs. the RV and BIS data sets.  
One observational epoch for \starD was removed because
we lack data for BIS.}
\begin{tabular}{
  l |
  S[table-format=1.2] | S[table-format=1.2] | S[table-format=1.2] | S[table-format=1.2] | 
  S[table-format=1.2] | S[table-format=1.2] | S[table-format=1.2] | S[table-format=1.2] | 
  r }
\hline
\multirow{3}{*}{Star}  & \multicolumn{8}{c}{TNG/HARPS-N} &   \\
  &\multicolumn{4}{c|}{$\ha$} &\multicolumn{4}{c}{$I_{\mathrm{Fe}}$} &  \\
  &\multicolumn{2}{c}{RV} &\multicolumn{2}{c|}{BIS}&\multicolumn{2}{c}{RV} &\multicolumn{2}{c}{BIS} &no \\
  &r &p&r &p&r &p&r &p \\
  \hline
  \starG &  0.17 & 0.49 & -0.40 & 0.10 & 0.29 & 0.25 & -0.20 & 0.42 & 18 \\ 
  \starA  &  0.02 & 0.93 & -0.11 & 0.70 &  0.23 & 0.40 & 0.27 & 0.31 & 16 \\ 
  \starF  &  0.10 & 0.68 &  0.04 & 0.87 & -0.03 & 0.91 & 0.73 & 0.00 & 18 \\ 
  \starH & -0.04 & 0.89 & -0.06 & 0.85 & 0.25 & 0.39 & -0.04 & 0.89 & 19 \\ 
  \starM &  0.55 & 0.03 &  0.61 & 0.02 &  0.53 & 0.04 &  0.76 & 0.00 & 15 \\ 
  \starC &  0.37 & 0.13 & 0.68 & 0.00 & 0.25 & 0.31 &  0.71 & 0.00 & 18 \\ 
  \starD &  0.03 & 0.90 & 0.00 & 0.99 &  0.74 & 0.00&   0.00 & 1.00 & 22\\ 
  \starE & -0.79 & 0.00 & -0.44 & 0.06 & 0.11 & 0.67 &  0.29 & 0.23 & 19\\ 
  \starI & -0.79 & 0.00 &  0.27 & 0.43 &  0.56 & 0.07 & -0.26 & 0.43 & 11\\ 
  \starJ &  0.31 & 0.25 & -0.53 & 0.03 & -0.02 & 0.93 & -0.06 & 0.84& 16\\ 
  \starK &  0.12 & 0.71 & -0.56 & 0.06 & -0.31 & 0.33 & -0.07 & 0.82 & 12\\ 
  \starL &  0.29 &0.39 & -0.29 &0.39 & -0.07 &0.84&  0.44 &0.18& 11\\ 
  \hline
  \end{tabular}
  \label{iha_tng}
\end{table*}

\subsection{Calcium H \& K doublet}

 The reversal profile in the cores of Ca H and K lines, that
is, the emission structure 
 at the core of the Ca absorption line, indicates increased stellar activity \citep{EberhardSchwarzschild1913}. 
The Ca II H \& K lines are not available from the HET/HRS spectra, but they are located 
at the blue end of the TNG/HARPS-N spectra. This is the region with the lowest S/N of all HARPS-N echelle orders.
The highest S/N for spectra or stars discussed in this paper is about 30, but it varies between 4 and 10 in general.
Stacking spectra to achieve a better S/N is not possible here as the spectra were taken at least a month apart, which is the adopted observing strategy in the TAPAS project. 
We manually examined Ca H \& K lines in every obtained TNG/HARPS-N spectrum and removed all epochs for which
stellar spectrum in the Ca region was dominated by instrumental response from the Ca analysis. These were spectra
with the lowest S/N, usually $<2$. 

None of single-epoch
observations for any star considered in this paper shows a reversal profile in the H \& K line cores. 

For every epoch (spectrum) of given star, we also calculated the $\shk$ index following 
\cite{Duncan1991}, and we calibrated it to Mount Wilson scale with the formula provided in \cite{2011arXiv1107.5325L}. 
The $\shk$ quantifies stellar chromospheric activity by comparing the flux in the cores of the H \& K
lines and two bands outside the lines.
The average $\shk$ values and their scatter for all epochs for a given star are presented in Table \ref{shk};
 they are typical for giant stars \citep{Zhao2013}. 

 Investigating the connection between RVs and $\shk$  is challenging for the available data set.
 First, the low S/N may affect its results; second, the
number of points for the statistics is quite low; and third, the observation span for TNG/HARPS-N data
 is shorter than three years. If the magnetic cycle is longer than that, we may therefore miss it 
 (i.e., the case of $\epsilon$Tau discussed in \citealt{Auriere2015}).
 To check whether the measured RVs are connected with the $\shk$ indexes, 
 we calculated the linear correlation coefficients between the
RVs and $\shk$.
As the blue end of the HARPS-N spectrum has a low S/N,
 we  calculated the correlation between the $\shk$ index and
the S/N to test if the S/N might be responsible for potential correlations.
 The results of this analysis are presented in  Table \ref{shk}.

We  found a statistically significant correlation between the
RVs 
and the $\shk$ activity indicator for one star - \starL. This means that for this star, the variability  in the RVs 
is associated with stellar activity.
For two other objects, \starA and \starF, the analysis may not be reliable
because there is a weak, but statistically significant correlation between $\shk$ and the S/N.

\begin{table*}
\centering
\caption{Data on $\shk$ activity.  The $r_{\mathrm{RV}}$ and $r_{\mathrm{S/N}}$
  are the Pearson correlation coefficients between RV and SHK, and between S/N and
  SHK, respectively. p is the significance level for the two-tailed test. }

\begin{tabular}{
    l | S[table-format=1.2] S[table-format=1.2] | S[table-format=1.2] S[table-format=1.2]  |
    S[table-format=1.2]  S[table-format=1.2] | S[table-format=1.0]}
\hline
  Star  & $\shk$  & $\sigma$  &$r_{\mathrm{RV}}$ & p &
  $r_{\mathrm{S/N}}$ &p&No  \\
\hline \hline
\starG & 0.17 & 0.02 & -0.37 & 0.15 &  0.42 & 0.09 & 17\\ 
\starA & 0.17 & 0.03 & -0.33 & 0.23 &  0.54 & 0.04 & 15 \\ 
\starF & 0.15 & 0.02 & -0.02 & 0.92 & -0.63 & 0.01 & 18\\ 
\starH & 0.17 & 0.02 & 0.17  & 0.58 & -0.29 & 0.34 & 13\\
\starM & 0.11 & 0.01 & -0.04 & 0.90 &  0.35 & 0.21 & 14 \\
\starC & 0.09 & 0.02 & -0.12 & 0.67 & 0.07 & 0.79 &16 \\ 
\starD & 0.13 & 0.03 & 0.05 & 0.84 & -0.32 & 0.14 & 22\\ 
\starE & 0.12 & 0.00 & -0.11 & 0.65 & 0.00 & 0.99 & 19\\ 
\starI & 0.16 & 0.03 & -0.41 &0.28 &  0.38 &0.32 & 9\\ 
\starJ & 0.10 & 0.03 & -0.25 &0.39 & -0.04 & 0.88& 14\\ 
\starK & 0.10 & 0.01 & -0.56 & 0.08 &0.47 & 0.15 & 11\\ 
\starL & 0.21 & 0.03 & -0.89 & 0.00 & -0.33 &0.32& 11\\ 
\hline
\end{tabular} 
\label{shk}
\end{table*}

\subsection{Photometry}

Stellar activity can  also manifest itself through changes in brightness. All our targets have been observed in
large photometric surveys. We collected available data for them from several different catalogs: ASAS \citep{ASAS}, NSVS \citep{NSVS},
Hipparcos \citep{Hipparcos}, and SuperWASP \citep{superWASP}. 
Table \ref{photo} presents the summary of the collected data. In data sets with many observing epochs 
(a star was observed multiple times in a one-day period), the data were resampled to one-day bins.
For stars that are included in multiple catalogs, we considered the longest set or the set with the most points to be the most reliable.
We searched for signals in the Lomb-Scargle periodogram for which the false-alarm probability (FAP) is greater than $10^{-3}$.
The photometric data sets present a measurable signal in the periodograms for only five stars. 

 \begin{table*}
\centering
\caption{Photometry collected from different catalogs for the
stars discussed in this paper. When the strength of the signal in the Lomb-Scargle periodogram exceeds $\mathrm{FAP} = 10^{-3}$, 
the corresponding change periods and  signal power
are provided.}
   \begin{tabular}{l | c S[table-format=4] S[table-format=4] S[table-format=2.3(4)] 
                   S[table-format=3.2] S[table-format=2.2] }
\hline
     Star & Catalog & {No. of points} & {Time range [d]} & {Av. bright. [mag]} &
     {period [d]} &  {Power}  \\
\hline \hline
\multirow{2}{*}{\starG} 
&ASAS & 250 &1840 & 9.014 \pm 0.014 & 29.5 & 16.82 \\ 
&NSVS & 26   &309 & 8.993 \pm 0.031 & {-} & {-} \\ 
\hline
\multirow{2}{*}{\starA} 
     & Hipparcos & 143 & 1124 & 8.854 \pm 0.017 & {-} & {-} \\
& SuperWASP &2948 & 1257 & 9.094 \pm 0.047 & 118.00& 16.02\\   
\hline     

\multirow{2}{*}{\starF} 
     &ASAS & 403  & 3157 & 8.031 \pm 0.012 & {-} & {-} \\
     &NSVS & 50   & 196  & 8.055 \pm 0.054 & {-} & {-} \\
\hline
\multirow{2}{*}{\starH} 
& ASAS & 354 &2648 & 8.469 \pm 0.012 & 1471.1 & 15.52\\ 
     &NSVS & 65   &196  & 8.455 \pm 0.029 & {-} & {-}\\
\hline
\multirow{3}{*}{\starM} 
     & Hipparcos & 130 & 1113 & 8.938 \pm 0.023 & {-} & {-} \\
     &NSVS & 82   &312    & 8.690 \pm 0.018 & {-}&{-}\\
     &SuperWASP &4597&458 & 9.155 \pm 0.019 & {-}&{-}\\
\hline
\multirow{2}{*}{\starC} 
     & ASAS & 388 & 3222& 10.263 \pm 0.017 &{-}&{-}\\
&NSVS &160& 204& 10.259 \pm 0.018 &44.3 & 12.14\\
\hline
     \starD & NSVS & 53 & 347& 9.774 \pm 0.023 &{-}&{-}\\
\hline
\starE 
     & Hipparcos & 106 & 1169 & 7.020 \pm 0.006 & {-}&{-}\\
\hline
\multirow{2}{*}{\starI} 
     &NSVS & 86   &216   &  9.408 \pm 0.020 & {-}&{-}\\
     &SuperWASP &5593&1475 & 9.877 \pm 0.024 & {-}&{-}\\
\hline 
\multirow{2}{*}{\starJ} 
     & NSVS & 100   &215   & 9.667 \pm 0.021 &{-}&{-}\\
     & SuperWASP & 5058 & 1262 & 10.137 \pm 0.011 &{-}&{-}\\
\hline
\multirow{2}{*}{\starK } 
     & NSVS & 90   &195   & 9.629 \pm 0.030 &{-}&{-}\\
     & SuperWASP &2114&600 & 9.839 \pm 0.011 & {-}&{-}\\
\hline
\multirow{2}{*}{\starL} 
     & NSVS & 217   &356  & 10.283 \pm 0.016 & {-}&{-}\\
& SuperWASP & 9055 & 458 & 10.589 \pm 0.023 & 27.91 & 15.61\\
 \hline
\end{tabular}
\label{photo}
\end{table*}

\section{Keplerian analysis \label{results-g}}

To determine the orbital parameters, we combined the global genetic algorithm (GA; \citealt{Charbonneau1995}) 
with the MPFit algorithm for the fitting \citep{Markwardt2009}. This hybrid approach  was described by \cite{Gozdziewski2003, Gozdziewski2006, Gozdziewski2007}.
The  range of Keplerian orbital parameters found with the GA was searched with the
RVLIN \citep{WrightHoward2009} code, which we modified to introduce stellar  jitter as a free parameter for fitting \citep{FordGregory2007, Johnson2011} the optimal solution.
The uncertainties were estimated using the bootstrapping method \citep{Murdoch1993, Kuerster1997, Marcy2005,Wright2007}.

For a more detailed description of the Keplerian analysis that
is briefly presented in this section, we refer to the first TAPAS paper, \cite{TAPAS1}.

\section{Lithium, beryllium, and carbon abundances and carbon isotopic ratios}

The Li abundances for 14 stars were derived in \cite{Adamow2014} and \cite{TAPAS2}. We used the same
method to calculate the Li abundance of \starF.

The carbon isotopic ratio $\ccr$ is an indicator of the evolutionary status of a star. To determine it, we combined all available TNG/HARPS-N spectra for a given star with an S/N higher than 30, 
which resulted in one spectrum with a high S/N and high resolution. In a first step, we analyzed two $\mathrm{C}_2$ bands in two spectral ranges,
$5150-5171\AA$ and $5625-5637\AA,$ to calculate the carbon abundance.
After setting the carbon abundance, we determined the $\ccr$ using the CH band at 
$4228-4240\AA$. These two steps of the analysis were made with the {\it synth} driver, which is part of the MOOG\footnote{
http://www.as.utexas.edu/$\sim$chris/moog.html} stellar line analysis code \citep{Sneden1973}. 
The S/N values for the combined spectra, the obtained carbon abundances, and for $\ccr$ are presented in Table \ref{carbon}.
 
Low beryllium  abundances for stars on the RGB are expected because
of the depletion at the first dredge-up episode and because non-standard
mixing processes start after the dredge-up phase. A more detailed description of A(Be) evolution along the RGB can be found in  \cite{TakedaTajitsu2014}.

Beryllium abundances were determined using MOOG synthetic spectrum fitting driver in the $3130-3133\AA$ spectral range. The analysis included 
two Be II lines at $3130.4$ and $3131.06\AA$. 
For giant stars, these lines are located in a crowded spectral region and are heavily blended,
mostly with iron and CH and OH bands.
Because of this, the continuum level is difficult to define. 
To obtain a proper continuum, we rescaled the observed spectrum around the Be lines so that it fitted the synthetic model as well as possible.  
The adopted carbon abundances and $\ccr$ ratios we obtained are presented in Table \ref{carbon}. 
Oxygen abundances come from \cite{Adamow2014}, that is, [O/H]= -0.13
and [O/H]= 0.17 for \starF and \starH, respectively. We adopted the solar value for beryllium $A(\mathrm{Be}_{\odot})=1.38$ \citep{Asplund2009}.
The best fit of the synthetic spectrum in the Be spectral region is provided using $A(\mathrm{Be})=-0.12$ for \starF and 
$A(\mathrm{Be})=-0.62$ for \starH.

\begin{table}
\centering
\caption{Carbon analysis. Adopted solar value: $A(\mathrm{C})_{\odot}=8.43$.} 
  \begin{tabular}{l | S[table-format=2] S[table-format=3] | S[table-format=1.2] S[table-format=2]}
\hline
    {Star} & {No.} & {S/N} & {[C/H]} & {$\ccr$}\\
\hline \hline
\starG	& 17 & 223 & -0.25 & 12 \\ 
\starA	& 16 & 259 & -0.30 &  18 \\ 
\starF	& 18 & 305 &  -0.25 &  20\\ 
\starH	& 14 & 283 &  -0.10&  10\\
\starM 	& 16 & 348 &  -0.25 & 20\\
\starC	& 17 & 203 & -0.15 & 12 \\ 
\starD	& 21 & 185 & -0.27 & 12 \\ 
\starE	& 19 & 454 & -0.20 & 10  \\ 
\starI		& 11 & 139 & -0.35 & 15 \\ 
\starJ	& 15 & 205 & -0.25 & 15  \\ 
\starK	& 12 & 194 & -0.40 & 23 \\ 
\starL	& 16 & 348 & -0.30&  17\\ 
\hline
\end{tabular}
\label{carbon}
\end{table}

   \begin{figure}
   \centering
   \includegraphics[width=\hsize]{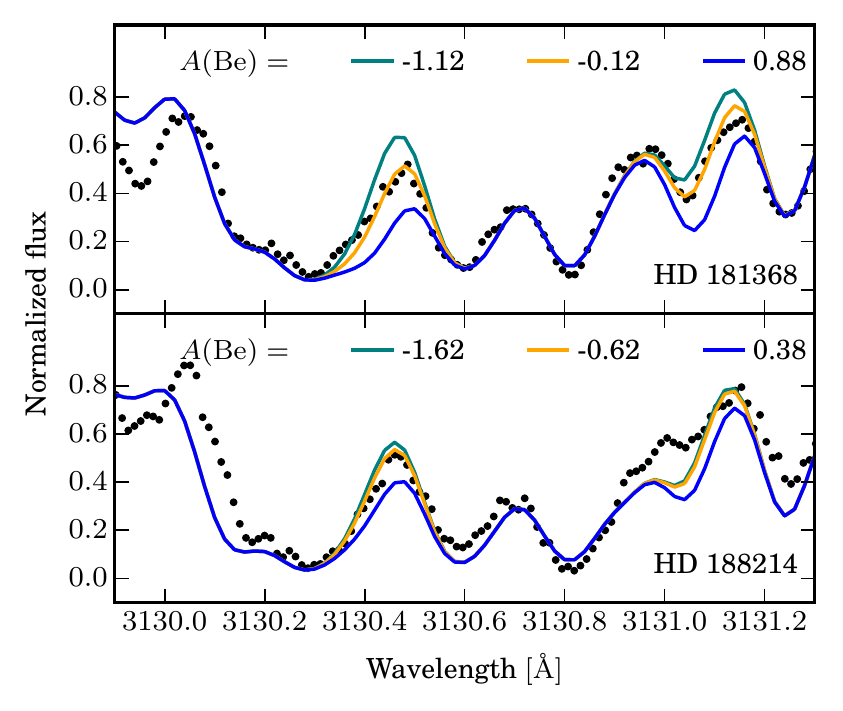}
      \caption{Results of the synthetic spectrum fitting to the
blue range of the UVES spectra. Black points represent the observed spectrum,  and solid lines are the synthetic spectra representing different Be abundances.}
         \label{be}
   \end{figure}

\section{Results}

The basic parameters of the stars studied in this paper are presented in Table \ref{parameters}.
Table \ref{RV-summary} delivers a short summary of radial velocity measurements and line bisector analysis.
In Tables \ref{iha_het},\ref{iha_tng} and \ref{shk} we present a summary of the spectral activity indicators, $\ha$ and $\shk$ indexes. 
An overview of the Fourier analysis for the available photometry of the program stars is presented in Table \ref{photo}.
Finally, a summary of the $\ccr$  analysis is presented in Table \ref{carbon}. To illustrate the current evolutionary status
of all our Li-rich giants (supported by results of our $\ccr$ analysis) we present them 
together with the complete PTPS sample of stars in H-R diagram in Figure \ref{HR}.

\subsection{Stellar activity indicators}

We investigated four indicators of stellar activity to check if it  is a likely source of observed RV variations. 
Based on collected RVs and stellar spectra we identified three active stars:
\starL (correlation between RV and $\shk$),
\starE (RV vs. BIS and RV vs. $\ha$ correlations)
and BD+02 3495 (RV vs. BIS correlation).

A comparison of statistically significant periods detected in the available photometry
for our targets stars (with false alarm probability $0.001\gtrsim \mathrm{FAP}$) is
presented in Table \ref{photo} with our estimates of the maximum rotation period derived
 from the estimated rotational velocity presented in Table \ref{parameters} 
shows that the available data allow for periodic stellar surface activity may be related to rotation (spots) in the case of two stars:
\starA, \starC.
In the case of \starH the period detected in ASAS photometry is way
too long to be connected with the rotation period of less than 76 days.
In the case of \starG and \starL the periods present in the photometric data 
are very similar to the lunar period from which they most likely originate.

\subsection{Li-rich giants with stellar companions}

In addition to TYC~0405-01700-1 and TYC~3314-01371-1 presented in \cite{Adamow2014} we found one more Li-rich giant with a stellar companion.
\starF, originally classified as a subgiant in PTPS sample \cite{Niedzielski2008}, was found by \cite{Deka2017} to be actually a giant.

We collected 27 epochs of RV data (9 from HET/HRS and 18 from
TNG/HARPS) between MJD= 54257 and  57238.  
There are no signs of stellar activity for this star, either in photometric data or line bisectors, that could 
explain the high amplitude of the RV variations (almost $5\kms$), hence we check the hypothesis that 
\starF hosts a companion. The strongest signal in the periodogram for RVs refers to the orbital period
of 1753 days, less than two times shorter that the time period covered with observations. 
Keplerian analysis of this signal led to solution that indicates that \starF hosts a low mass stellar companion. 
But after removing the Keplerian fit from the RV data, we got a strong, periodic signal ($P\sim1000$ days) in RV residuals, that could be 
interpreted as an additional planetary mass companion. Both objects would orbit \starF on  very eccentric orbits ($e>0.8$)
what might raise questions about the long term stability of such a system. 
Another solution is to assume that the signal observed in the periodogram is caused by the fact that the cadence of the observations 
is short compared to the orbital period, hence the Lomb-Scargle analysis signal in RVs may not be reliable.

During a second approach to Keplerian fitting we chose to explore
the parameter space with longer orbital periods. Results of that analysis are presented in Table \ref{KeplerianFit}
and the RV measurements and the fit are depicted in Fig. \ref{Fit_1509}.
As a result, we got an orbital fit with low RMS and no additional signal in residua. 
The companion is a low mass star with $m\sin i = 0.4\Msun$ on a very elliptical, long period orbit. The semi-major axis
of 14.8~au would place it between orbits of Saturn and Uranus in Solar System.

\starF has a typical spectrum of a red giant. We did not observe any distortion in the shape of CCF functions 
that could prove that the companion contributes to observed spectrum. 
This star is one of two objects we measured Be abundance. Obtained $A(\mathrm{Be}) = -0.12$ is typical
for a star at this evolutionary stage. Therefore, for \starF Li enhancement is not associated with elevated
Be abundance.

\begin{table*}
\centering
  \caption{Keplerian orbital parameters of the companions to the stars discussed in this paper. }
\renewcommand{\arraystretch}{1.3}
\begin{tabular}{lllllll}
\hline  
  Parameter               &  \scriptsize  \starF b        & \scriptsize \starA b      & \scriptsize  \starM b     & \scriptsize  \starJ b        & \scriptsize  \starK b       & \scriptsize \starD b     \\
\hline
\hline
\normalsize
  $P$ (days)              & $12977.6^{+0.3}_{-0.35}$     & $733^{+5}_{-8}$          & $4100^{+210}_{-240}$     & $562^{+4.1}_{-4.0}$         & $130.48^{+0.51}_{-0.70}$   & $339.5^{+1.3}_{-2.3}$   \\
  $T_0$ (MJD)             & $69449^{+0.3}_{-0.34}$       & $54774^{+21}_{-15}$      & $60540^{+230}_{-250}$    & $54340^{+60}_{-60}$         & $ 54066^{+13}_{-12}$       & $54927^{+10}_{-6}$      \\
  $K$ (\!\ms)             & $2947^{+3.3}_{-3.4}$         & $54^{+5.2}_{-8}$         & $71^{+7}_{-11}$          & $75.42^{+0.90}_{-0.7}$      & $30^{+5}_{-2.7}$           & $190^{+40}_{-120}$      \\
  $e$                     & $0.7524^{+0.0004}_{-0.0004}$ & $0.76^{+0.05}_{-0.09}$   & $0.56^{+0.07}_{-0.05}$   & $0.098^{+0.082}_{-0.032}$   & $0.48^{+0.2}_{-0.12}$      & $0.78^{+0.03}_{-0.23}$  \\
  $\omega$ (deg)          & $293.9^{+0.1}_{-0.1}$        & $100^{+12}_{-15}$        & $318^{+8}_{-7}$          & $129^{+40}_{-36}$           & $14^{+20}_{-23}$           & $22^{+25}_{-7}$         \\
  $m_2\sin i$ (\!\Mjup)   & $220 \pm 21$                 & $1.7 \pm 0.7$            & $6.0 \pm 2.7$            & $3.42 \pm 0.35$             & $1.33 \pm 0.44$            & $4.6 \pm 5.2$           \\
  $a$ (\!\au)             & $10.7 \pm 0.5$               & $1.7 \pm 0.1$            & $5.7 \pm 0.9$            & $1.414 \pm 0.063$           & $0.72 \pm 0.02$            & $1.02 \pm 0.03$         \\
  $V_0$ (\!\ms)           & $-55868^{+4}_{-4}$           & $-45932.5^{+0.7}_{-1.1}$ & $-6724^{+14}_{-8}$       & $24810^{+210}_{-170}$       & $19314.7^{+3.1}_{-1.2}$    & $-50145^{+4}_{-17}$     \\
  $V_1$ (\!\ms / day)     & -                            & -                        & -                        & $-0.102^{+0.003}_{-0.004}$  & -                          & -                       \\
  offset (\!\ms)          & $56906^{+8}_{-8}$            & $45924^{+8}_{-8}$        & $6743^{+18}_{-21}$       & $-19143^{+9}_{-10}$         & $19314^{+7}_{-7}$          & $50180^{+9}_{-8}$       \\ 
  \sjit (\!\ms)           & $4.0^{+0.5}_{-0.8}$          &  $17.8$                  & $14.5$                   & $11.6$                      & $5.77$                     & $15$                    \\
  $\sqrt{\chi_\nu^2}$     & $1.36$                       & $1.24$                   & $1.28$                   & $1.31$                      & $1.58$                     & 1.29                    \\
  RMS (\!\ms)             & $6.3$                        & $21.7$                   & $19.8$                   & $15$                        & $13.2$                     & 18.7                    \\
  $N_{\textrm{obs}}$      & $27$                         & 41                       & $86$                     & $28$                        & $32$                       & 36                      \\
\hline
\end{tabular}
\renewcommand{\arraystretch}{1}
\tablefoot{$V_0$ denotes the absolute velocity of the barycenter of the system,
the offset is a shift in RV measurements between different telescopes,
\sjit~is the stellar intrinsic jitter as defined in \cite{Johnson2011},
RMS~is the root mean square of the residuals.}
\label{KeplerianFit}
\end{table*}

\begin{figure}
   \centering
   \includegraphics[width=0.5\textwidth]{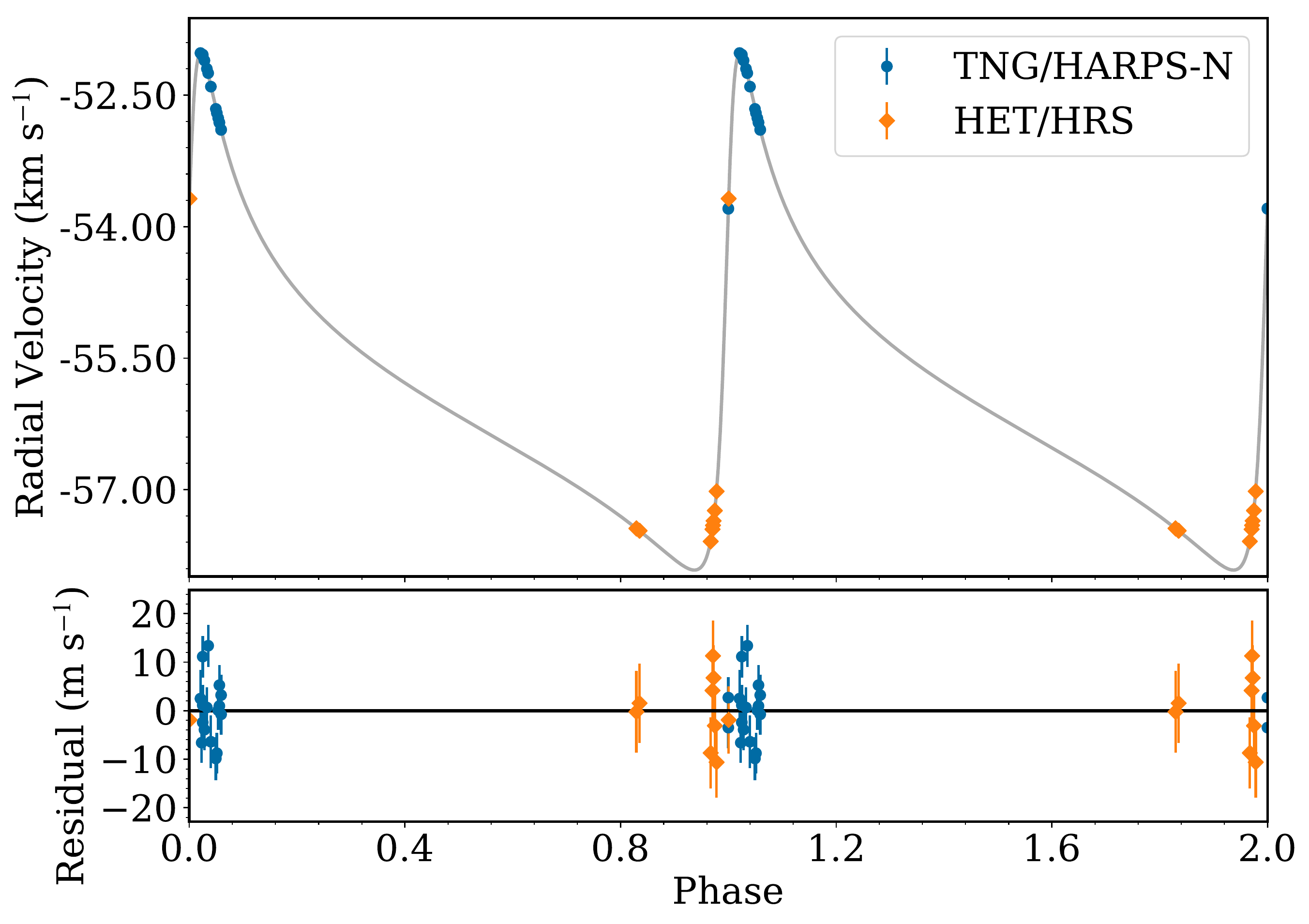}
   \caption{Keplerian best fit to the combined HET/HRS and TNG/HARPS-N data for
      \starF. The jitter is added to the uncertainties.}
   \label{Fit_1509}
\end{figure}

\subsection{Li-rich giants with planetary-mass companions}

We report here two new candidates for planetary-mass companions to Li-rich giants and present a confirmation and updated parameters for \starA .

\subsubsection{\starA  : updated orbit} 

This is a $M/\Msun =1.12$ , $R/\Rsun=19.27$ \citep{Adamczyk2016}, $\log g = 2.48$  giant star.
It was the first Li-rich giant reported as a potential planetary host by \cite{Adamow2012} 
and a possible case of Li enrichment by engulfment of close-in planets.

The low carbon isotopic ratio of $\ccr=18$ obtained by us indicates that \starA\; completed FDU depletion and is a regular giant on the RGB.

In addition to the 15 epochs of precise RV presented in \cite{Adamow2014}, we collected 11 new RV measurements from HET/HRS 
and 16 new RV measurements with HARPS-N under the TAPAS project
for this star.
In the extended set of observations, the Keplerian hypothesis concerning the apparent RV variations is still valid: 
no RV vs. line bisector correlation was found, and no indication of stellar activity can be extracted from existing data.
The periodic signal in SuperWASP photometry may indicate a spot rotating with a star, but because of the very different period,
it is unrelated to the observed Keplerian RV signal.

The updated orbital solutions for this star are based on the total number of 41 RV 
measurements obtained over 3491 days, between MJD = 53747.2 and MJD=57238.2.
New orbital parameters for this system are presented in Table \ref{KeplerianFit}, and the best Keplerian fit is shown in Fig. \ref{Fit_1015}. 
The updated  set of orbital parameters is very similar to the original set. 
We therefore confirm our previous finding that this star is a good candidate for a planetary
host with a companion with a minimum mass of $1.7\Mjup$ on a very elliptical  orbit ($e=0.76$).

\begin{figure}
   \centering
   \includegraphics[width=0.5\textwidth]{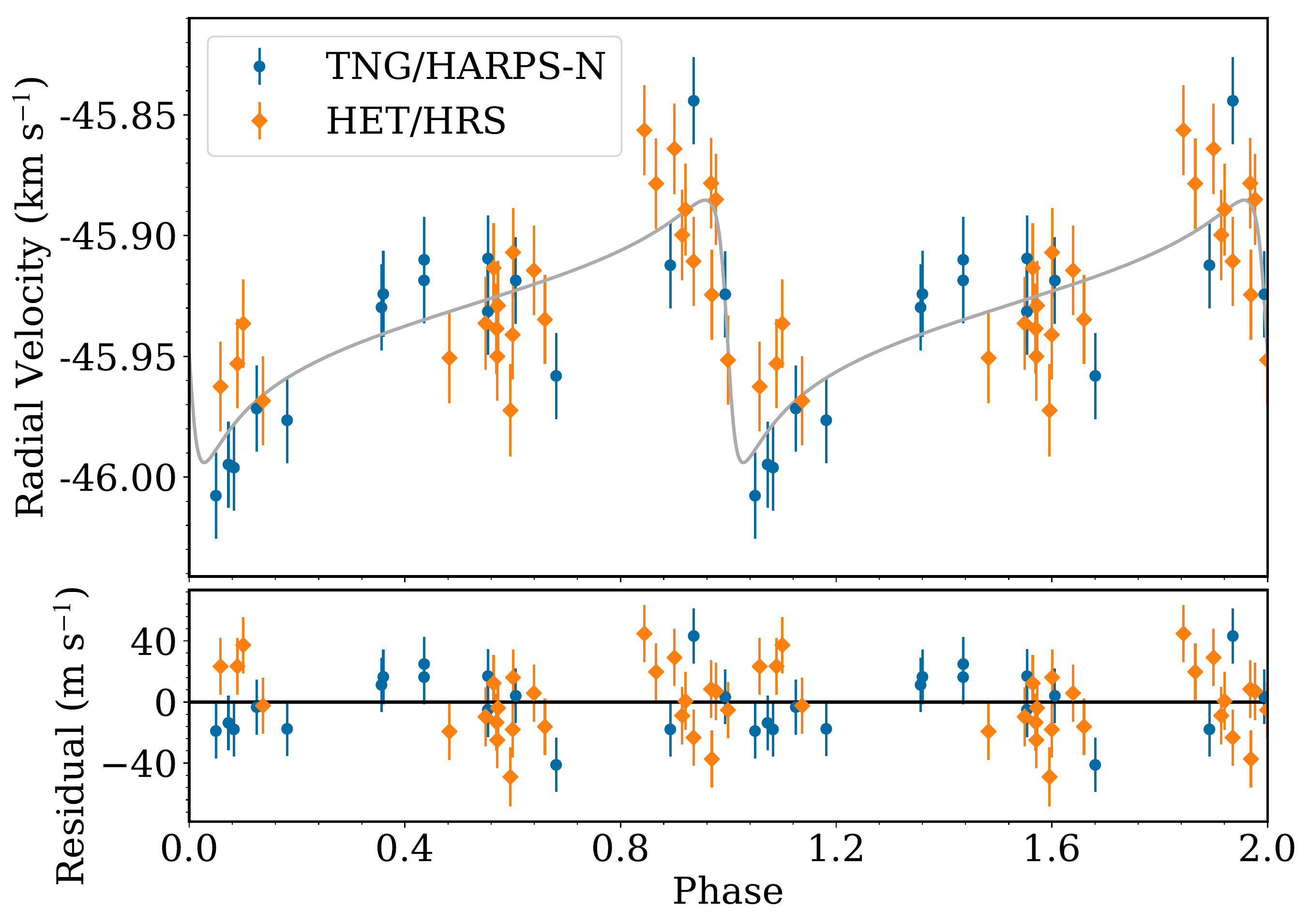}
   \caption{Keplerian best fit to the combined HET/HRS and TNG/HARPS-N data for
      \starA. The jitter is added to the uncertainties.}
   \label{Fit_1015}
\end{figure}

\subsubsection{\starM:  a giant planet in 5.7 au orbit} 

The source \starM  (BD+59 1909, TYC~3917-01107-1)  is a $V=8.79\pm0.01$ \citep{Tycho2},   $\pi=1.70\pm0.23$ \citep{TGAS}  star in Draco.
Atmospheric parameters derived by \cite{Zielinski2012} are presented in Table \ref{parameters} 
together with physical parameters derived  by \cite{Deka2017}.
This star is a giant ($\log g = 2.37$) with a mass of  
$M/\Msun =1.47$, radius  $R/\Rsun=12.73$ \citep{Adamczyk2016} and $[\mathrm{Fe}/\mathrm{H}] = -0.25$.
The advanced evolutionary stage is confirmed with a low $\ccr$ ratio.

We collected a total number of 86 epochs of RV measurements for this object between MJD = 53294.1 and MJD = 57238.1,
spanning the time period of 3944 days. 
No indication of stellar activity was found in  HET/HRS or TNG/HARPS-N data,
 which allowed for a Keplerian interpretation of the apparent RV variations.
The observed RV changes presented in Fig. \ref{Fit_1254} can be explained by the presence of a  $6\Mjup$ companion on an eccentric orbit with $e=0.56$.
The orbital period  is 4100 days (Table \ref{KeplerianFit}).

\begin{figure}
   \centering
   \includegraphics[width=0.5\textwidth]{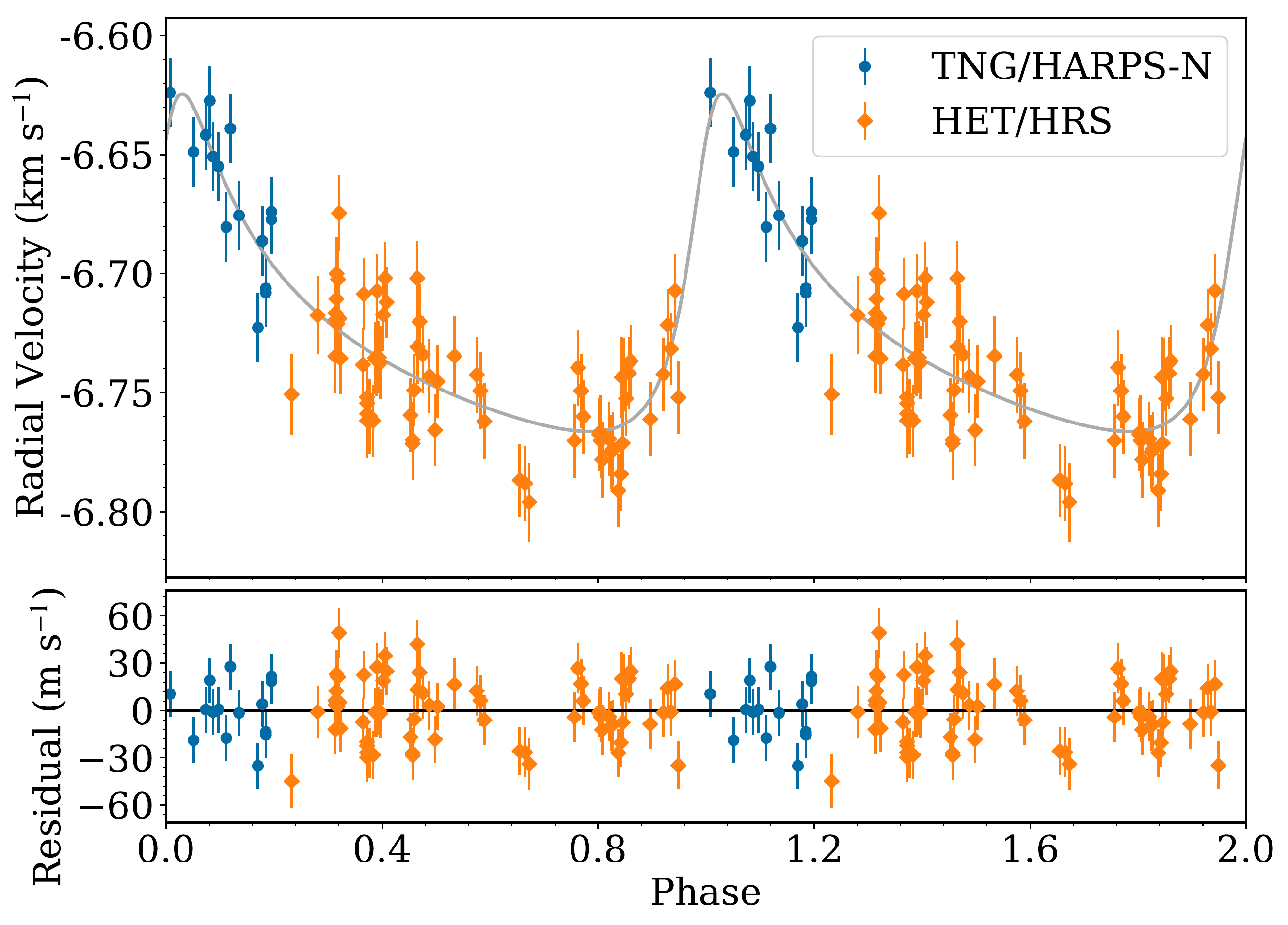}
   \caption{Keplerian best fit to the combined HET/HRS and TNG/HARPS-N data for
      \starM. The jitter is added to the uncertainties.}
   \label{Fit_1254}
\end{figure}

\subsubsection{\starJ: a planet in a binary system}\label{PTPS1029} 

The source \starJ (Gaia 439327806662223744, 2MASS J03021978+4944389 ) is a $V=9.90\pm0.03$ \citep{Tycho2} and $\pi=1.80\pm0.39$ \citep{TGAS}
giant star with $\Teff = 4776$ K, $\log g = 2.97$, and solar metallicity ($[\mathrm{Fe/H}] = -0.06$) in Perseus. According to \cite{Deka2017}, it has a mass of 
$M/\Msun=1.19$ and a radius of $R/\Rsun=5.90$.
With $\ccr = 15$ and $\ALi =1.51,$ this star was classified as
a Li-rich giant after the FDU stage.

A total number of 28 epochs (12 from HET/HRS and 16 from TNG/HARPS-N) of RV measurements were collected for this object
over 3514 days, between MJD = 53724.2 and MJD=57238.2. 
No correlation between RV and BIS is present in either HET/HRS or TNG/Harps data.
The obtained data show a linear trend in RV vs. MJD (Fig. \ref{Fit_1029}, upper panel) and periodic changes in RV,
which may be interpreted as a presence of a low-mass companion (Fig. \ref{Fit_1029}, lower panel). 
The orbital solution fitted 
to detrended data is presented in Table \ref{KeplerianFit}. \starJ is a potential host to a $3.42\Mjup$ planet on a 562-days slightly 
elliptical ($e=0.098$) orbit. 

After removing the fitted RV trend that is due to a distant companion, we checked the activity indicators again 
and found a statistically significant correlation between 
 $\ha$ and RV (r=0.75, p=0.01) in the HET/HRS data. 
 Since a similar correlation appears between $\ha$ and RV in the GC flat spectra (r=0.72, p=0.01)
 and no trace of any activity was found in TNG/HARPS-N data, we assume that this stems solely from weak iodine lines. 
For the detrended TNG/HARPS-N data, there is no correlation for $\ha$ vs. RV (r=0.01,p=98 for $\ha$, r=0.08, p=78 for $\ife$),
and neither does the $\shk$ index show a correlation with RV (r=0.33, p=0.26).
Consequently, there is no compelling evidence of a stellar activity nature of the observed RV variations for \starJ.

\begin{figure}
   \centering
   \includegraphics[width=0.5\textwidth]{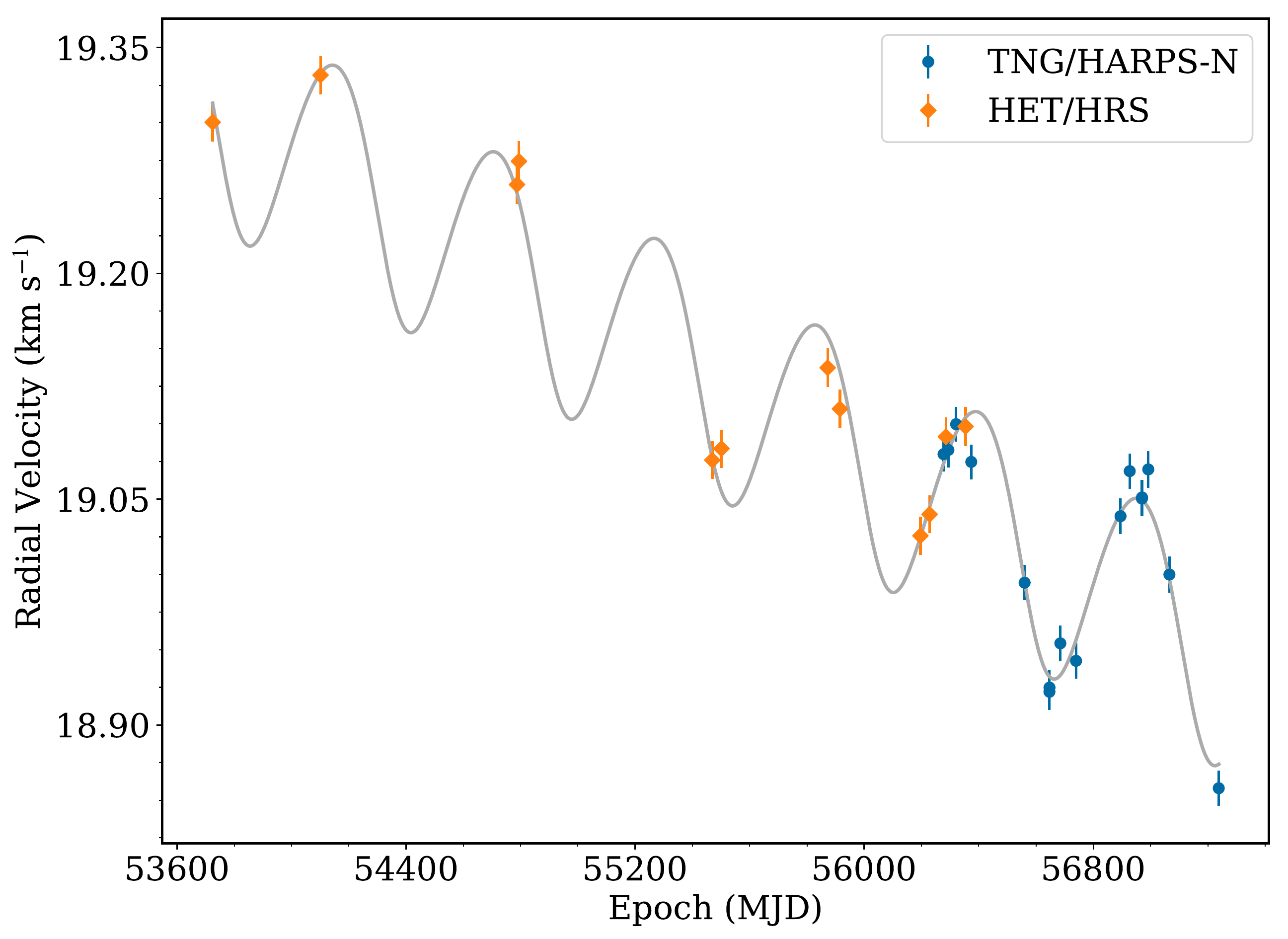}
   \includegraphics[width=0.5\textwidth]{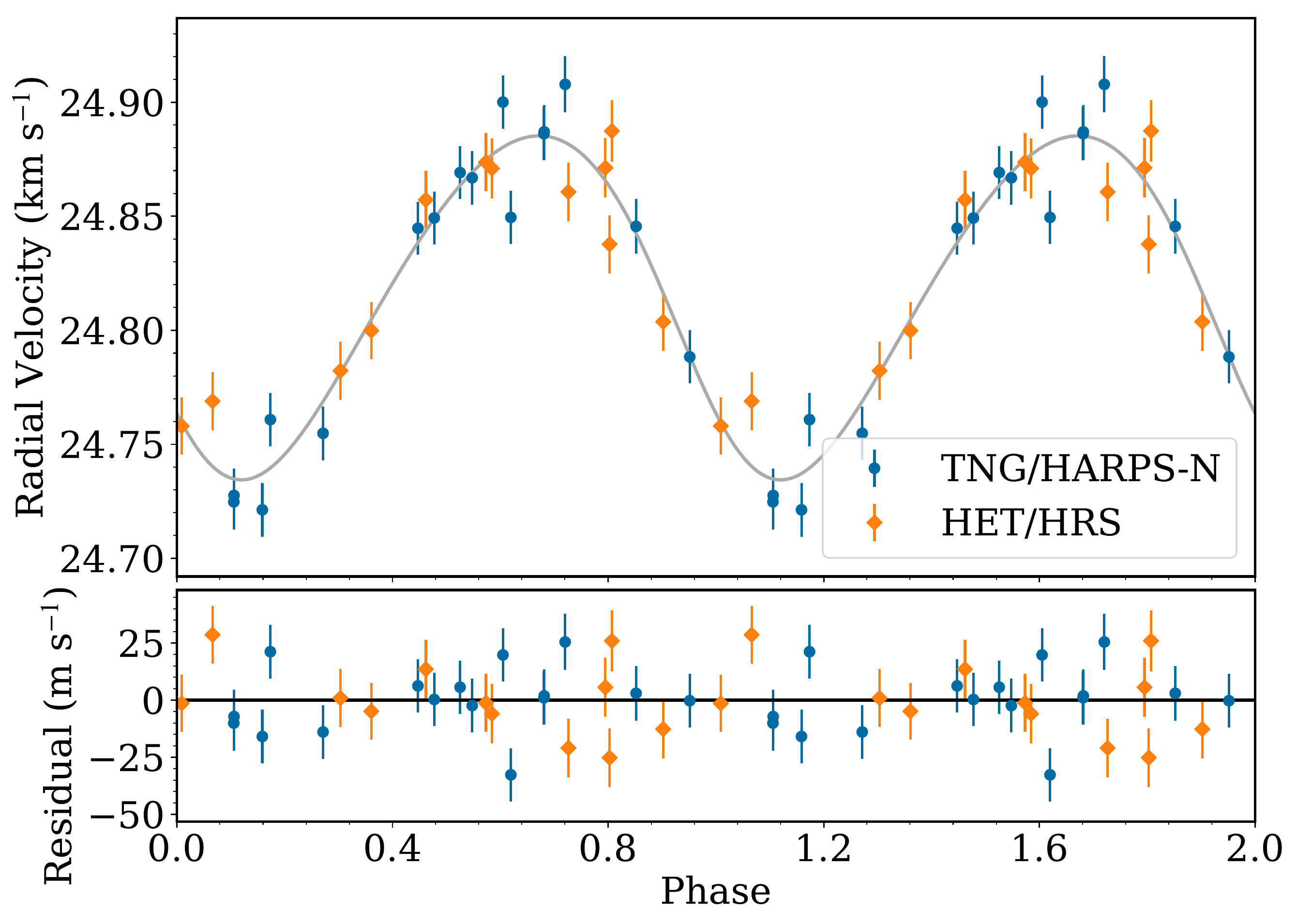} 
   \caption{Keplerian best fit to the combined HET/HRS and TNG/HARPS-N data for
      \starJ. The jitter is added to the uncertainties.}
   \label{Fit_1029}
\end{figure}

\subsection{Possible companions}

For two of the stars, \starK and \starD, we collected enough data to perform the Keplerian analysis, but since the phase coverage 
is not complete, they require confirmation with future observations.

\subsubsection{\starK: a low-mass planet at the edge of engulfment} 

This star is a giant with following parameters: $\Teff=5068$~K, $\log g=2.40$, $[\mathrm{Fe/H}]=-0.26$, $M/\Msun = 2.88$, and $R/\Rsun = 17.66$.
We collected 32 epochs of RV measurements for this object, 20 from HET/HTS and 12 from TNG/HARPS-N. 
The observed RV variations are at least 11 times larger than the RV uncertainty and at least 4.6 times larger than the expected RV amplitude of the p-mode oscillations. 
The star is clearly RV variable and not only due to oscillations. There is no observational evidence of stellar 
activity in this star, so that the observed RV variations may be caused by Doppler shifts that are due to a companion.
The observed variations seem to be periodic, with an amplitude of $30\ms$, hence we performed a Keplerian fit to the data. 
The RV curve we obtained is presented in Fig. \ref{Fit_1162}, and the parameters of the best fit are included in
Table~\ref{KeplerianFit}. 
We found \starK to be the possible host of a $0.85\Mjup$, close-in ($a=0.57$~au, $P=130$~days) planet on an elliptical orbit ($e=0.48$).

\begin{figure}
   \centering
   \includegraphics[width=0.5\textwidth]{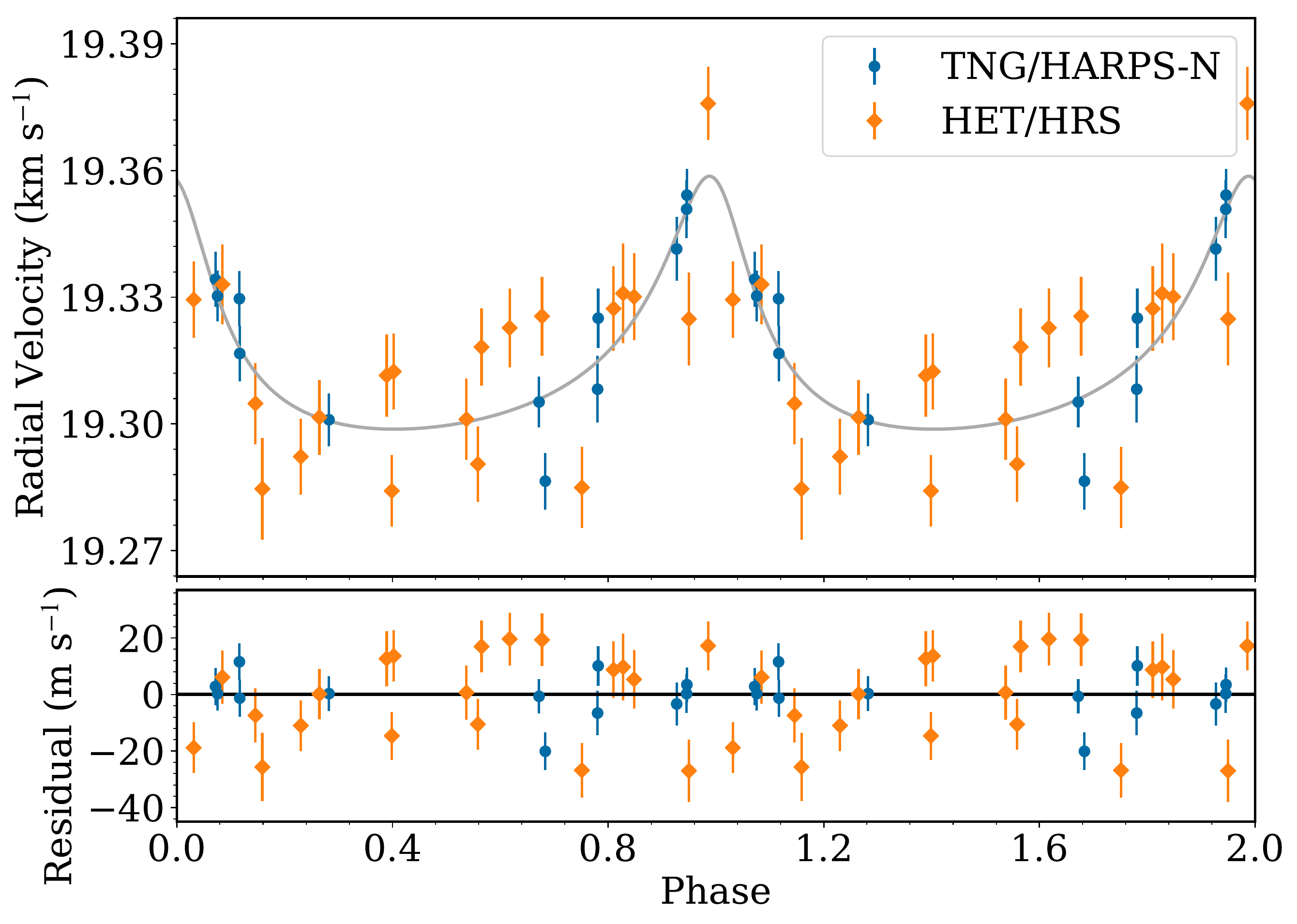}
   \caption{Keplerian best fit to the combined HET HRS and TNG Harps-N data for
      \starK. The jitter is added to the uncertainties.}
   \label{Fit_1162}
\end{figure}

\subsubsection{\starD: a planet on a highly eccentric orbit} 

The source \starD has $\Teff = 4673$~K, $\log g = 2.45$, $[\mathrm{Fe/H}] = -0.14$, $M/\Msun =1.22 $, and $R/\Rsun=10.62,$ and it is another 
case of a star that may host a low-mass companion on a very eccentric orbit.
For this object we collected 14 HET/HRS observational data points within the PTPS project
and an additional 16 epochs form TNG/HARPS-N follow-up. 

The observed RV amplitude for \starD is 24 times larger than the RV uncertainty and 14 times larger 
than the expected RV amplitude of the p-mode oscillations.
The data collected here show no trace of stellar activity in \starD, 
 so that the only viable source of the variability seems to be a companion. 

The obtained radial velocity curve is presented
in Fig. \ref{Fit_0926}, and details of the fitted Keplerian solution are listed in Table~\ref{KeplerianFit}. 
A $4.9\Mjup$ companion might orbit the star on a 339.5-day orbit
with $e=0.781$.

\begin{figure}
   \centering
   \includegraphics[width=0.5\textwidth]{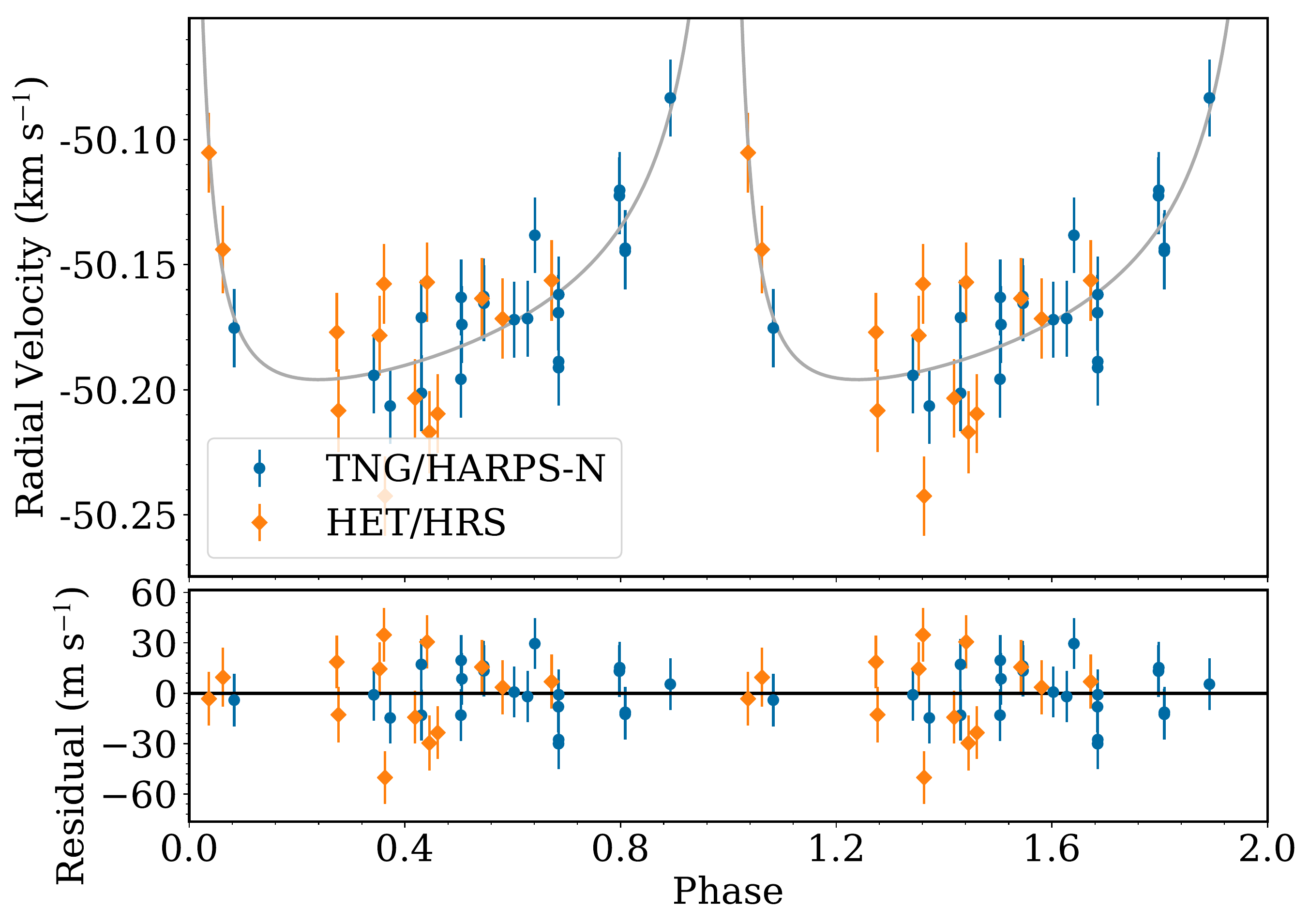}
   \caption{Keplerian best fit to the combined HET HRS and TNG Harps-N data for
      \starD. The jitter is added to the uncertainties.}
   \label{Fit_0926}
\end{figure}

\subsection{Single or unresolved stars}

Seven stars in our sample of Li-rich giants are objects that
we classified as single.

The most interesting of them is \starB, the most Li-rich giant in our sample, a star described in detail in \cite{TAPAS2}. 
This object is a very interesting case 
for Li studies, however, because it is a star on the early RGB phase and Li production could not have been triggered 
by the mechanisms that have been proposed for enrichment in giants as they occur in a later evolutionary stage.

Another two stars in this group are \starH and \starL, for which the estimated p-mode 
oscillation RV amplitudes in (sparse) HET/HRS data are comparable to the RV amplitude. 
Although the more numerous HARPS-N data show larger RV amplitudes, no significant periodic variations were detected. 

\starL shows a correlation between the observed HARPS-N RV variations and the $\shk$ index. 
We also detected a strong periodic signal in the available SuperWASP photometry for this object, but with
a period of 27.91 days it is most likely attributable to lunar phases. 
We consider this star as single and active. 

\starH shows no trace of stellar activity in the spectra that
were available to us. It does show 
a strong photometric period of 1471.1 days in the ASAS data,
however. This may by physical, 
although it is apparently unconnected with the much slower rotation period of 71~days.
The enhancement of Li does not occur simultaneously 
with the Be enrichment - $A(\mathrm{Be})$ of -0.62 for \starH is low,
and typical for giants at this evolutionary stage.

For the remaining four objects 
\starG, 
\starC, 
\starE, and
\starI,
we collected a sufficient number of observations to obtain conclusive orbital solutions, 
but it was not possible to identify any periodic signals in their RVs data that might have 
been generated by a low-mass companion.
Three of them, \starC, \starG, and \starE, show correlations between the RV and activity indicators, 
hence we consider them active.

\section{Discussion} 

 Our in-depth study of 12 of the 15 most Li abundant giants from the PTPS sample \citep{Adamow2014} 
resulted in the discovery of two new candidates for planetary systems around Li-rich giants and in the confirmation of one that was presented 
in \cite{Adamow2012}. We also found one new stellar binary and two stars that appear to have planetary-mass companions.
Another seven stars were found to be single. 

Of special interest are two of the newly discovered planetary hosts, or host candidates, presented here.
\starM b is a unique planet, the most distant that orbits a giant star known so far ($a =5.7$~au).
It may only be compared to HD~14067 b \citep{Wang2014}, an object of higher mass ($9\Mjup$) 
that does not reside that far away from its star ($a=5.3$~au). Owing to its high eccentricity ($e=0.56$), 
\starM b
cruises between 2.5 and 8.9 au, leaving little space for other planets in this system.
At this orbital separation, the planet is likely to survive the evolution of its $1.47\Msun$ host at least up to the AGB phase.

Even more intriguing is the possible planetary companion of \starK, an evolved ($\log g =2.40$), 
extended ($17.66\Rsun$), very rare intermediate-mass (2.88 M$_{\odot}$) star. Owing to its proximity to the star 
and to the high orbital eccentricity, its relatively low-mass ($1.33\Mjup$) companion that we advocate 
for may periodically approach the red giant at only 4.5 stellar radii, which makes it potentially 
a perfect laboratory to study tidal interactions before it is
eventually ingested by its sun. 
This potential planetary system certainly deserves more attention.

It is also worth mentioning that the updated orbit of \starA b confirms that this case has the most eccentric orbit ($e=0.76$) 
of a planet around a giant star (a record that may be beaten by \starD, if confirmed with future observations).

\subsection{Planet frequency around Li-rich giants in the PTPS sample}

Altogether, the sample of the 15 most Li-rich giants in the PTPS sample contains 
three planetary-mass companions (one in a binary system), 
two possible planetary-mass companions, 
three binary systems, 
and seven single stars.
 In other words, $53\%$ of the stars in the PTPS Li-rich giant sample appear to have companions. 
The planet occurrence rate is therefore at least $20\%$ and may even be 
as high as $33\%$. The binary star frequency is $27\%$. Forty-seven percent of the PTPS Li-rich objects appear to be single.
The frequency of RV-variable field red giants was estimated to
be $30\%$ by 
\cite{GunnGriffin1979}.
\cite{MermilliodMayor1980} found $15\%$ of the giants in open clusters to be RV variable. 
\cite{HarrisMcClure1983} found $15-20 \%$ of the cluster giants to be RV variable.
\cite{MermilliodMayor1989} found $25-33\%$ of the red giants in clusters to be RV variable.
The frequency of RV-variable stars (binaries) in our sample appears to be consistent with these results.

The gas-giant planet fraction around evolved solar-mass stars with lower than solar metallicity 
was recently estimated by \cite{Reffert2015} 
as $6.9-11.1\%$. The frequency of planets around the 15 most Li-rich giants in our sample 
is therefore about three times higher.
Our results are more consistent with the results presented by these authors 
for giants with higher than solar metallicity ($25\%-28.6\%$). The problem here 
is apparently the different sensitivity of our survey and that of \cite{Reffert2015}. 
These authors assumed that they are sensitive to companion masses of $2.3\Mjup$ with a period up to 2.3 years.

When we assume that the minimum mass of a companion detectable in our search is about $1\Mjup$ 
and the maximum orbital separation is about 6 au, our results are more consistent with 
the frequency of planets around dwarfs presented by \cite{Cumming2008}, 
that is, $17\%-20\%$ of the stars with gas giants with orbital periods $\gtrsim 300$ d, typical for planets around giants.

However, the relatively high frequency of planetary-mass companions 
around Li-rich giants suggests a relation between the Li overabundance in giants and the presence of exoplanets.

\subsection{Nature of eccentricity}

Except for one object (\starJ), all identified candidates for planetary
hosts have planets on highly eccentric orbits with eccentricities
of between 0.48 and 0.76,
well above average for planets around evolved stars. The orbits of stellar companions
around Li-rich giants appear to be more circular ($e=0.04-0.339$, \citealt{TAPAS2}), 
except for \starF with its $e=0.7524$.

Of other planets around Li-rich giants, orbital elements are available only for two: 
8~UMi (HD~133086) \citep{Kumar2011, Lee2015} and NGC~2423~3 \citep{Carlberg2016}. 
Contrary to the planets presented in this paper, they both reside on more circular orbits with $e=0.21$ 
and $e=0.06$, respectively.

The engulfment of inner planets, and as a consequence, the disruption of a planetary system, 
is a good explanation for the high eccentricities as well as
the Li enrichment.
\cite{AguileraGomez2016} calculated that an engulfment episode can explain a Li enrichment of up to $\ALi =2.2$, 
while higher abundances require other enrichment mechanisms. All but one of the potential hosts with low-mass companions discussed here 
have $\ALi<2.2,$ hence the engulfment scenario is feasible. \starD is the exception with $\ALi_{NLTE} = 2.86$.

A strong indication of such a process might come from infrared excess,
which should be associated with Li enhancement through engulfment episodes \citep{SiessLivio1999}. 
The connection between Li enhancement and infrared excess was studied in \cite{Adamow2014} and 
\cite{Rebull2015}. None the objects discussed here shows any excess in infrared photometry.

\subsection{Connection to stellar clusters?}

High eccentricity is rather uncommon in evolved planetary systems around giants, although the most recent discoveries of high-eccentricity systems 
(see, e.g., \citealt{Witten2017} and the stars reported in this paper) seem to be reversing this trend.
For main-sequence systems, high eccentricity has been suggested
to imply a violent dynamical history of the observed systems
or a Kozai resonance with a distant companion \citep{TakedaRasio2005}. In search for a scenario 
leading to such a configuration, we note that the most suitable environment in which 
dynamical interactions may produce planetary systems with a highly eccentric component 
and in which an enhanced probability of planet engulfment through star-planet collision may be witnessed
are open clusters. According to simulations of the dynamical evolution of an open cluster 
member star hosting planets by \cite{Hao2013}, up to 12$\%$ 
of the inner planets in their equal-planetary mass models may eventually collide with the stars. 
The outer planets gain high eccentricity in the process. Similar results were more recently
 obtained by \cite{Shara2016} in simulations of two-planet systems in open clusters. 
 Star-planet collisions are only the extreme case of dynamical interactions between 
 a multi-planet system and other open cluster members. The most common results 
 in the simulations of the above mentioned authors are hot Jupiters in very tight orbits. 
 We note that this process, together with stellar evolution that
naturally leads to an ingestion 
 of the inner planets and to star-planet collisions, creates ideal conditions for the pollution 
 of the stellar surface with elements from one or more planets, possibly leading to an enhanced 
 Li surface abundance and an enhanced eccentricity of the surviving planets.

We therefore checked whether our Li-rich giants with planets were members of a cluster, for which we used data on open clusters from
\cite{Kharchenko2013}.
We found that three stars from our sample are located within three cluster radii ($R_{\mathrm{C}}$) from open cluster centers: 
\starG (Collinder 359), \starI ($\alpha$ Per), and \starJ ($\alpha$ Per), the last star within $1 R_{\mathrm{C}}$. 
In the case of \starJ, we also found matching metallicities. 
However, neither RV nor proper motions match, which makes a cluster membership for these stars unlikely.

The simulations of \cite{Villaver2014} predict 
a modest decay of the semi-major axis and eccentricity before a planet is engulfed. 
This means that a high-eccentricity planet can sustain the RGB evolution at high eccentricity.

\section{Conclusions  \label{conclusions}}
 
 We presented two new candidates for planetary systems around Li-rich giants: 
 \starM with an $m_{p}\sin{i}=6.0 \Mjup$ companion in $a=5.7$ au, $e=0.56$ orbit, 
 and \starJ, a component of a binary system with a $m_{p}\sin{i}=3.42 \Mjup$ companion 
 in $a=1.4$~au, $e=0.09$ orbit. We have updated the orbital parameters 
 for \starA and also found a binary Li-rich giant, \starF, with an $m\sin{i} = 220 \Mjup$ 
 companion in $a=10.7$ au, $e=0.7524$ orbit.
 In the analysis of another two Li-rich giants presented in this paper, 
 \starK and \starD, we argue for planetary-mass companions in eccentric orbits, but we lack the data to prove it.
 
 One out of three binary systems and two (four) out of three (five) planetary-mass companions 
 orbit their hosts on eccentric orbits with $e>0.48$. We note, however, that the other two Li-rich giant planets with available orbital elements, 
 8~UMi (HD~133086) \citep{Kumar2011, Lee2015} and NGC~2423~3 \citep{Carlberg2016},
 show more circular orbits ($e\leq0.21$).
 
 In our sample of the 15 most Li-rich giants from the PTPS, 
 we find three planetary-mass companions, two possible planetary-mass companions, 
 three binary systems and one binary with a planet, and seven single stars. The binary frequency of $27\%$ appears 
to be normal, but the $20-33\%$ planetary companions frequency is a factor of 3 higher than
in the general sample of giants with planets \citep{Reffert2015}.
 
The high planet occurrence rate for Li-rich giants presented here suggests a relation between Li-abundance and planets from
 Li-rich giants and the pollution of the stellar surface with Li from a low-mass companion as the Li enhancement mechanism. 
 The high eccentricity in the majority of the planetary systems we presented suggests previous dynamical interactions 
 in these systems either with other planets within the system or with more distant, stellar companions. 
 We therefore investigated the possibility of planetary system disruption through Kozai resonances
 in open clusters, but we found that none of our stars is a member of an open cluster.
 Neither did we find any trace of infrared excess, which is assumed to be associated with the engulfment process \citep{delaRezaDrake1997}.
 
 We found no Be surface abundance anomalies in the two Li-rich giants \starF and \starH.
 These two stars have low Be abundances, typical for their evolutionary stage.
 Although it is difficult to generalize these two estimates to all our stars, they clearly do not 
 provide evidence of a planetary engulfment process in these stars.
 
 The evolution of planetary systems, the main mechanisms of which are summarized by \cite{Veras2016},
 is a complex process that is influenced by many factors. 
 The Li-rich planetary hosts discussed here represent an interesting opportunity to study this problem in more detail.

\begin{acknowledgements}
 
We would like to thank the referee, whose comments helped improve and clarify this manuscript.
We also thank the HET, IAC, and ESO resident astronomers and telescope operators for
their support.

MA acknowledges the Mobility+III fellowship from the Polish Ministry of Science
and Higher Education. 

AN was supported by the Polish National Science Centre
grant no. UMO-2015/19/B/ST9/02937.

EV acknowledges support from the Spanish Ministerio de Econom\'ia y Competitividad under grant AYA2013-45347P.
KK was funded in part by the Gordon and Betty Moore Foundation's
Data-Driven Discovery Initiative through Grant GBMF4561.

This research was supported in part by PL-Grid Infrastructure.

The HET is a
joint project of the University of Texas at Austin, the Pennsylvania State
University, Stanford University, Ludwig-Maximilians-Universit\"at M\"unchen,
and Georg-August-Universit\"at G\"ottingen. The HET is named in honor of its
principal benefactors, William P. Hobby and Robert E. Eberly. 
The Center for Exoplanets and Habitable Worlds is supported by the Pennsylvania State
University, the Eberly College of Science, and the Pennsylvania Space Grant
Consortium.

This research has made use of the SIMBAD database, operated at CDS, Strasbourg, France. This research has made use of
NASA's Astrophysics Data System. The acknowledgements were compiled using the Astronomy Acknowledgement Generator.
This research made use of SciPy~\citep{jones_scipy_2001}. This research made use of the yt-project, a toolkit for
analyzing and visualizing quantitative data~\citep{Turk:2011}. This research made use of matplotlib, a Python library
for publication quality graphics~\citep{Hunter:2007}. This research made use of Astropy, a community-developed core
Python package for Astronomy~\citep{2013A&A...558A..33A}. IRAF is distributed by the National Optical Astronomy
Observatory, which is operated by the Association of Universities for Research in Astronomy (AURA) under cooperative
agreement with the National Science Foundation~\citep{1993ASPC...52..173T}. This research made use of
NumPy~\citep{van2011numpy}. 

\end{acknowledgements}


\bibliographystyle{aa} 

\end{document}